\pdfoutput=1
\documentclass[aps,reprint,superscriptaddress,nofootinbib,prl,preprintnumbers]{revtex4-2}
\usepackage{amsmath, amssymb, hyperref, color, graphicx, feynmp-auto}
\usepackage[normalem]{ulem}
\definecolor{nicered}{rgb}{.7,.1,.1}
\definecolor{nicegreen}{rgb}{.1,.5,.1}
\definecolor{darkblue}{rgb}{0,0,.5}
\definecolor{el}{rgb}{.9, .8, .7}
\definecolor{mu}{rgb}{.8, .7, .8}
\definecolor{lightgray}{gray}{0.6}
\hypersetup{colorlinks, citecolor=nicegreen, linkcolor=nicered, urlcolor=darkblue}

\usepackage{amsmath, amsfonts, slashed, graphicx, color, colortbl, hyperref, multirow}
\definecolor{nicered}{rgb}{.7,.1,.1}
\definecolor{nicegreen}{rgb}{.1,.5,.1}
\definecolor{darkblue}{rgb}{0,0,.5}
\hypersetup{colorlinks, citecolor=nicegreen, linkcolor=nicered, urlcolor=darkblue}

\definecolor{pnk}{rgb}{0.96, 0.69, 0.83}
\definecolor{lpnk}{rgb}{1.0, 0.89, 0.95}
\definecolor{dcyn}{rgb}{0.78, 0.92, 0.91}
\definecolor{cyn}{rgb}{0.80, 0.98, 0.95} 
\definecolor{grn}{rgb}{0.85, 0.96, 0.82} 
\definecolor{gld}{rgb}{0.98, 0.86, 0.58}
\definecolor{blu}{rgb}{0.78, 0.91, 1.0}
\definecolor{lgr}{rgb}{0.93, 0.93, 0.93}

\begin{document}

\title{Hadron Collider Signatures of Lepton Number Violation in the Type II Seesaw Model}

\author{Patrick D. Bolton}
\email{patrick.bolton@ijs.si}
\affiliation{Jo\v{z}ef Stefan Institute, Jamova 39, 1000 Ljubljana, Slovenia}

\author{Jonathan Kriewald}
\email{jonathan.kriewald@ijs.si}
\affiliation{Jo\v{z}ef Stefan Institute, Jamova 39, 1000 Ljubljana, Slovenia}

\author{Miha Nemev\v{s}ek}
\email{miha.nemevsek@ijs.si}
\affiliation{Faculty of Mathematics and Physics, University of Ljubljana, Jadranska 19, 1000 Ljubljana, Slovenia}
\affiliation{Jo\v{z}ef Stefan Institute, Jamova 39, 1000 Ljubljana, Slovenia}

\author{Fabrizio Nesti}
\email{fabrizio.nesti@aquila.infn.it}
\affiliation{\normalsize \it 
  Dipartimento di Scienze Fisiche e Chimiche, Universit\`a dell'Aquila, via Vetoio, I-67100, L'Aquila, Italy}
\affiliation{\normalsize \it INFN, Laboratori Nazionali del Gran Sasso, I-67100 Assergi (AQ), Italy}

\author{Juan Carlos Vasquez}
\email{jvasquezcarm@umass.edu}
\affiliation{
Department of Physics and Astronomy, Science Center, Amherst College, Amherst, MA 01002, USA.}

\date{\today}

\begin{abstract}
\noindent 
We examine the prospect of observing genuine lepton number violating (LNV) signals at hadron 
colliders in the context of the Type~II seesaw mechanism.
The model features smoking gun signals involving same-sign di-leptons and jets that may be the 
primary observable channel in certain regions of the parameter space.
The flavour composition of final-state charged leptons is related to the 
origin of neutrino masses and is correlated with other rare processes, such as neutrinoless 
double beta decay.
We review existing collider limits and provide sensitivity estimates from LNV signals at upcoming 
LHC runs, including non-zero mass splittings between triplet components.
\end{abstract}


\maketitle

\noindent {\bf Introduction.}
A well-known feature of the Standard Model (SM) is the accidental conservation of 
total lepton number $L$.
The observation of non-zero neutrino masses could imply the existence of 
$L$ number violating (LNV) new physics.
According to Majorana~\cite{Majorana:1937vz}, a real 
representation under the Lorentz group can describe neutrinos with a $\Delta L = 2$ 
mass term.
The most studied LNV process, sensitive to light neutrino 
exchange and new physics, is neutrinoless double beta decay~($0\nu\beta\beta$)~\cite{Racah:1937qq}.
Considerable experimental and theoretical efforts are underway to detect such smoking gun 
signals and establish its connection to light neutrinos~\cite{Vissani:1999tu}
and new physics~\cite{Tello:2010am}; for a review, see~\cite{Dolinski:2019nrj}.

Other LNV processes have been probed from eV to beyond TeV~\cite{Cai:2017mow}.
Perhaps the most striking signature is the final state of two same-sign charged leptons 
and two jets in $pp$ collisions, arising from the production and decay of 
heavy Majorana neutrinos $N$.
Such states are the key ingredient of Type~I seesaw scenarios~\cite{Minkowski:1977sc, 
Mohapatra:1979ia, Glashow:1979nm, GellMann:1980vs, Yanagida:1979as},
which can introduce Majorana masses $m_N$ for the gauge singlets $\nu_R$ either by hand 
or via the spontaneous symmetry breaking of an extended gauge group, such as in the minimal 
Left-Right symmetric model (LRSM)~\cite{Pati:1974yy, Mohapatra:1974hk, Senjanovic:1975rk, 
Mohapatra:1979ia}.
There, it is tied to the scale of $SU(2)_R$ breaking through the Yukawa coupling $Y_N$ as 
$m_N = Y_N v_R$, where $v_R$ is the vacuum expectation value (VEV) of the scalar triplet $\Delta_R$.

In models with gauged $U(1)_{B-L}$ (and $SU(2)_R$), the VEV $v_R$ can be in TeV 
range, kinematically accessible to colliders. 
Distinct LNV signatures, such as the Keung-Senjanović~\cite{Keung:1983uu} channel, appear nearly 
automatically due to the presence of $m_N$ and Yukawa couplings that communicate LNV to the SM.
The production and decay of on-shell $N$ via charged currents leads to LNV in half of events, 
containing two same-sign leptons and jets.
In the case of interference, the ratio of same- and opposite-sign leptons may be 
altered~\cite{Gluza:2016qqv} and LNV may be suppressed by pseudo-Dirac 
masses~\cite{Kersten:2007vk, Drewes:2019byd}.
The exact origin of heavy Majorana neutrino masses can be determined from complementary signals, 
such as the Higgs decay $h \to N N$ via the $h-\Delta_R^0$ 
mixing~\cite{Maiezza:2015lza, Nemevsek:2016enw}, which can also feature LNV final states 
with two same-sign leptons and up to four jets.
See~\cite{Nemevsek:2018bbt} for a review of the LRSM parameter space 
and~\cite{Nemevsek:2023hwx} for future colliders.

\begin{figure}
  \centering
  \input{pp_DD_VVll.tex}
  \vspace*{-1ex}
  \caption{
  Pair (and associated) production of charged scalars mediated by $Z/\gamma^*$ ($W^\pm$), 
  producing the LNV final state $\ell^{\pm}\ell^{\prime\pm} 4j$ at the LHC.}
  \label{figLNVchannels}
\end{figure}

The Type~II seesaw~\cite{Magg:1980ut, Schechter:1980gr, Cheng:1980qt, Mohapatra:1980yp, 
Lazarides:1980nt} provides Majorana masses for the light neutrinos without extending
the SM gauge group or fermion content.
A single scalar triplet $\Delta_L = (3,2)$ under $SU(2)_L \otimes U(1)_Y$ can partially break 
the electroweak (EW) symmetry with its VEV $v_\Delta$.
The Yukawa term that couples the left-handed lepton doublet to the triplet with the 
coupling $Y_\Delta$ then leads to the Majorana neutrino mass matrix 
$M_\nu \propto Y_\Delta v_\Delta$.

The production of charged and neutral scalars in the model proceeds through the EW interactions~\cite{
Han:2007bk, Fuks:2019clu}, followed by decays defining the final state~\cite{Chun:2003ej, 
Garayoa:2007fw, Kadastik:2007yd, Perez:2008ha, Das:2024yvt}.
For $v_\Delta \lesssim 10~\text{keV}$, leptonic decay modes dominate with three and 
four-lepton states, with current flavour-dependent lower bounds on the doubly charged scalar 
mass in the $\mathcal O(500-700)~\text{GeV}$ range~\cite{ATLAS:2017xqs, CMS:2012dun, 
L3:2003zst, ATLAS:2014vih, OPAL:2001luy, CMS-PAS-HIG-16-036, ATLAS:2014kca}.
In such final states, the total lepton number is conserved (LNC).
Another LNC search exists in the $v_\Delta \gtrsim 100~\text{keV}$ region, where decays 
to SM vector bosons take over, with weaker bounds at 
$m_{\Delta^{++}}\gtrsim 350 \text{ GeV}$~\cite{ATLAS:2021jol, ATLAS:2018ceg, ATLAS:2024itc}. 
Mass splittings between $\Delta_L$ components trigger LNC cascade 
decays~\cite{Melfo:2011nx, Primulando:2019evb} that either enhance or reduce the bounds.
Signatures of Type~II have been investigated at future lepton~\cite{Agrawal:2018pci}, 
$e-p$~\cite{Dev:2019hev} and hadron~\cite{Du:2018eaw, Padhan:2019jlc} colliders.

It is well-known~\cite{Maiezza:2016bqj} that genuine LNV signals in Type II require the 
simultaneous presence of $Y_\Delta$ and $v_\Delta$.
Turning on the Yukawa coupling $Y_\Delta$ ensures that the $\Delta_L$ triplet indeed 
has $L = 2$.
Then, $v_\Delta$ communicates $L \neq 0$ to the SM sector with $L = 0$ and the combined 
effect leads to LNV.
This is especially obvious when the final-state gauge bosons decay hadronically and 
neutrinos cannot carry away $L$.
In pure Type~II, the product $Y_\Delta v_\Delta$ is fixed and LNV signals are not 
automatically observable.

\begin{figure}[t!]
  \centering
  \includegraphics[width=.90\columnwidth]{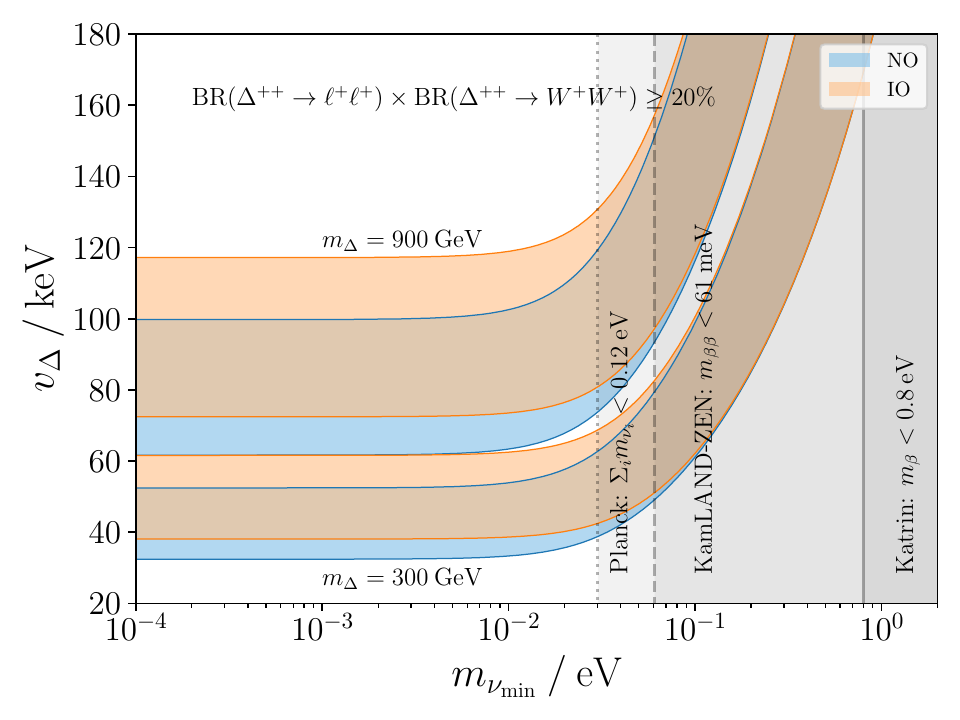}%
\vspace*{-2ex}%
\caption{Position of the LNV window in the Type~II seesaw, defined by 
  $\text{BR}_{\Delta^{++} \to \ell^+ \ell^+} \times \text{BR}_{\Delta^{++} \to W^+ W^+} > 20\%$.
  Blue and orange bands correspond to the NO and IO cases, each with the benchmark masses $m_{\Delta^{++}} = 300~\text{GeV}$ and $900~\text{GeV}$.
  The gray regions indicate the limits from Planck data~\cite{planck:2018vyg} 
  (dotted vertical line), $0\nu\beta\beta$ from KamLAND-Zen~\cite{KamLAND-Zen:2016pfg} 
  (dashed) and KATRIN~\cite{KATRIN:2021uub} (solid).}
  \label{figLNVWin}
\end{figure}

The LNV window~\cite{Maiezza:2016bqj} exists for $v_\Delta \simeq 10 - 100 \text{ keV}$, 
where both decay modes, proportional to $Y_\Delta$ and $v_\Delta$, are present.
Smoking gun signals with two same-sign leptons and jets appear, shown in FIG.~\ref{figLNVchannels}.
While existing leptonic searches can be recast to the 
LNV region~\cite{delAguila:2013mia, Babu:2022ycv} with some sensitivity, a dedicated analysis 
presented here opens the LNV window completely, and accounts for the presence of mass splittings
in $\Delta_L$.
We determine the strength of the LNV signal with a di-lepton plus 
jets analysis that significantly improves the sensitivity, spanning the range of 
$v_\Delta$ shown in~FIG.~\ref{figLNVWin}.

\medskip\noindent {\bf Location of the LNV window.}
In the Type~II seesaw, the Yukawa term
\begin{align} \label{eq:Yukawa}
\mathcal{L}_{\text{Yuk}} \supset - {Y_\Delta}_{ij} L_i^T \mathcal C i \sigma_2 \Delta_L L_j + \text{h.c.} \, ,
\end{align}
with
\begin{equation} 
\label{eq:triplet}
\Delta_L = 
	\begin{pmatrix} \Delta^+/\sqrt{2} & \Delta^{++} 
  	\\
  	\left(v_\Delta + \Delta^0 + i \chi_\Delta \right)/\sqrt 2 & - \Delta^+/\sqrt{2} 
  	\end{pmatrix} \, ,
\end{equation}
sources the light neutrino mass matrix
\begin{align} \label{eqMnu}
  M_\nu &= V^* m_\nu V^\dagger = \sqrt 2 Y_\Delta v_\Delta \, ,
\end{align}
where $m_\nu = \text{diag}(m_1,m_2,m_3)$ contains the light neutrino masses and $V$ is the PMNS 
mixing matrix. 
The singly charged $\Delta^+$ and pseudo-scalar $\chi_\Delta$ components mix with the SM 
would-be Goldstones, while the neutral component $\Delta^0$ mixes with the 
SM Higgs $h$ with a small mixing of order $v_\Delta/v$.
Eq.~\eqref{eq:Yukawa} also results in the decays of $\Delta^{++}$ to two same-sign charged 
leptons, with
\begin{align} \label{eqDpplilj}
  \Gamma_{\Delta^{++} \to \ell^+_i \ell^+_j}&= \frac{m_{\Delta^{++}}}{8 \pi \left(1 + \delta_{ij} \right)} 
  \left|\frac{M_{\nu i j}}{v_\Delta}\right|^2 \, ,
\end{align}
where $\delta_{ij}$ is the Kronecker delta. Through $M_{\nu}$, and hence $m_\nu$ and $V$, 
these rates are directly related to neutrino oscillations~\cite{Chun:2003ej, Garayoa:2007fw,
Kadastik:2007yd}. 
The total leptonic rate,
$\Gamma_{\Delta^{++} \to \ell^+ \ell^+} = m_{\Delta^{++}}/(16 \pi) \sum m_\nu^2/v_\Delta^2$,
is insensitive to $V$ and depends only on the neutrino masses, determined from the mass-squared 
differences $\Delta m_{21}^2$ and $\Delta m_{32}^2$ and the lightest neutrino mass $m_{\nu_{\min}}$.
With $m_\nu$ fixed by oscillations, the leptonic rates in~\eqref{eqDpplilj} dominate the 
$\Gamma_{\Delta^{++}}^{\text{tot}}$ when $v_\Delta$ becomes small, 
until the lower bound on $v_\Delta$ is met, coming from the non-observation of lepton flavour 
violating (LFV) processes~\cite{Abada:2007ux, Fukuyama:2009xk, Dev:2018sel, Barrie:2022ake, 
Ardu:2023yyw, Banerjee:2024lsi} and $Y_\Delta$ perturbativity.
For the larger values of $v_\Delta$ relevant for the LNV window, the LFV rates are highly suppressed
by small Yukawa couplings and the constraints become irrelevant.

\begin{figure}
  \centering
  \includegraphics[width=.90\columnwidth]{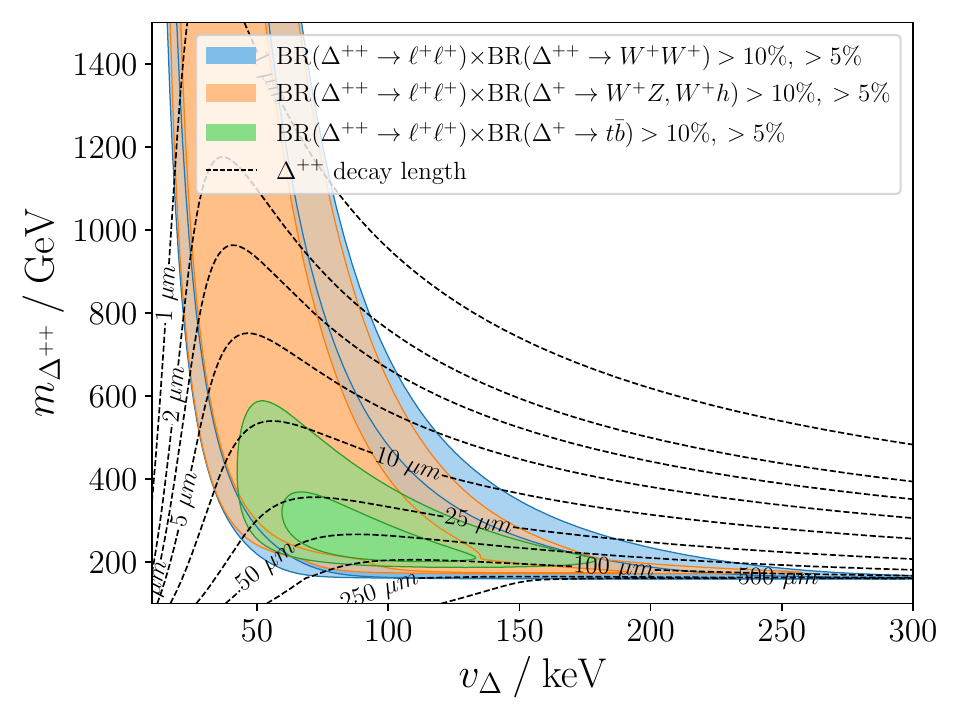}%
  \vspace*{-2ex}%
  \caption{Regions with the branching ratios of
  $\text{BR}_{\Delta^{\pm\pm}\rightarrow \ell^{\pm}\ell^{\pm}} \times
  \{\text{BR}_{\Delta^{\pm\pm}\rightarrow W^{\pm}W^{\pm}},
  \text{BR}_{\Delta^{\pm}\rightarrow W^{\pm}Z(h)},
  \text{BR}_{\Delta^{\pm} \rightarrow t \overline b}\} > 5\%, 10\%$ in blue, orange and green.
  The NO mass spectrum is assumed with $m_{\nu_{\rm min}} = 0.01 \text{ eV}$.
  The dashed lines show the $\Delta^{++}$ decay length in the rest frame,
  see~\cite{Dev:2018kpa, Antusch:2018svb, Alimena:2019zri, Arbelaez:2019cmj} for displacement.}
  \label{figLNVWin2}
\end{figure}

Likewise, $v_\Delta \neq 0$ triggers decays of $\Delta_L$ components into the SM gauge 
bosons via the kinetic term $\text{Tr}\left[(D _\mu \Delta_L)^\dagger (D^\mu \Delta_L)\right]$.
Particularly important are the decays $\Delta^{++} \to W^+W^+$ and $\Delta^{+} \to W^+ Z$, 
with the rates
\begin{align}
  \Gamma_{\{W^+ W^+, W^+ Z\}} &\simeq \frac{\alpha_2}{4} \left(\frac{v_\Delta}{v}\right)^2 
  \left \{ \frac{m_{\Delta^{++}}^3}{M_W^2}, \frac{m_{\Delta^{+}}^3}{2M_W^2} \right \} \, ,
\end{align}
for $m_{\Delta^{+}}, m_{\Delta^{++}} \gg v$.
Also relevant are $\Delta^+ \to W^+ h$ and $\Delta^+ \to t\bar{b}$ that proceed 
via $\Delta^+-\chi^+$ mixing, with 
\begin{align}
  \Gamma_{\Delta^+ \to W^+ h} &\simeq \frac{\alpha_2}{8} \left( \frac{v_\Delta}{v} \right)^2 \frac{m_{\Delta^+}^3}{M_W^2} \, ,
  \\
  \Gamma_{\Delta^+ \to t \bar b} &\simeq \frac{3}{4 \pi} \left( \frac{v_\Delta}{v} \right)^2 m_{\Delta^+} 
  \left( \frac{m_t}{v} \right)^2 \, .
\end{align}
Unlike~\eqref{eqDpplilj}, these are independent of $m_\nu$ and dominate for larger values 
of $v_\Delta$. 
The upper bound of $v_\Delta \lesssim \mathcal{O}(1)~\text{GeV}$ comes from electroweak 
precision tests (EWPT).

The number of genuine LNV signal events is proportional to the 
product of leptonic and gauge branching ratios,
$\text{BR}_{\Delta^{++} \to \ell \ell} \times \{ \text{BR}_{\Delta^{++} \to WW},\, 
\text{BR}_{\Delta^+ \to WZ(h)}, \,\text{BR}_{\Delta^+\to t \bar b} \}$
that define the LNV window.
Its position depends on the neutrino mass ordering but less so on $m_{\nu_{\min}}$,
non-trivial behaviour happens in the region disfavoured by the cosmological 
data~\cite{planck:2018vyg} and the latest $0\nu\beta\beta$~\cite{KamLAND-Zen:2016pfg}
and KATRIN~\cite{KATRIN:2021uub} searches.
The exact location is shown in FIGs.~\ref{figLNVWin} and~\ref{figLNVWin2} and depends 
non-trivially on $m_{\Delta^{++}}$, which we scan within the LHC observable range.
It is maximal for $v_\Delta \sim 40-50 \text{ keV}$.

\begin{figure}
  \centering
  \includegraphics[width=0.90\columnwidth]{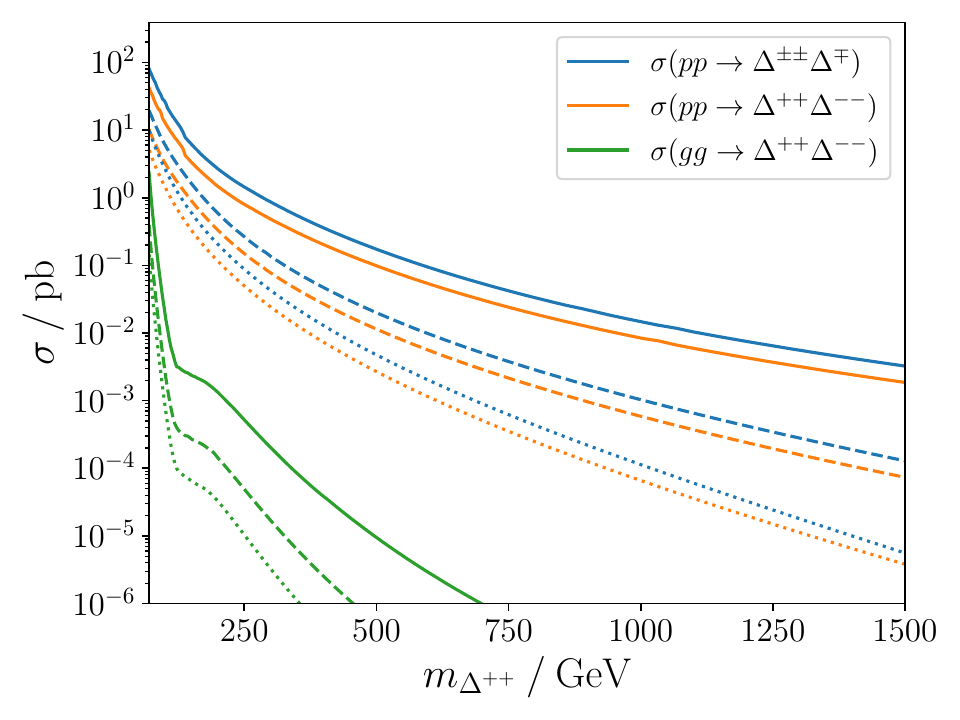}%
  \vspace*{-2ex}%
  \caption{Proton level cross-sections at $\sqrt{s} = 14,\,27,\,100~\mathrm{TeV}$ in
  dotted, dashed and solid lines with $\Delta m = 0$, for the LO associated, EW pair 
  and gluon fusion Higgs pair production.}
  \label{figXSecs}
\end{figure}

\medskip\noindent {\bf Size of the LNV window.}
To obtain the number of LNV events, the branching ratios above are multiplied by the dominant 
production cross-sections.
The doubly charged scalars can be pair-produced in $pp$ collisions through off-shell
$\gamma/Z/h \rightarrow\Delta^{\pm\pm}\Delta^{\mp\mp}$ and in the associated channel 
via $W^\pm \rightarrow\Delta^{\pm\pm}\Delta^{\mp}$, see FIG.~\ref{figLNVchannels}.
The associated production cross section is integrated over the parton distribution 
functions (PDFs) as
\begin{equation}
  \sigma^{\text{LO}}_{\text{assoc.}} \simeq \int_{\text{PDF}} \frac{\pi \alpha_2^2}{36} 
  \frac{\hat s  \left(1 - 4 m_{\Delta^{++}}^2 /{\hat s} \right)^{3/2}}{(\hat s - M_W)^2 
  + (\Gamma_W M_W)^2} \, ,
\end{equation}
and similarly for $\gamma/Z/h$ exchange. 
The leading order (LO) cross-sections at different $\sqrt s$, multiplied by the 
next-to-leading-order (NLO) $K$-factors, are shown in FIG.~\ref{figXSecs}.
We assumed $m_{\Delta^{++}} = m_{\Delta^+} = m_{\Delta^0}$, the impact of mass splittings
is discussed below.
For details of NLO cross-sections and PDF uncertainties, see~\cite{Muhlleitner:2003me, Fuks:2019clu}.
The kinematics of NLO jet emissions does not impact our study and the LO simulations 
are sufficient for an accurate estimate of signal selection efficiencies in the LNV window.

\medskip\noindent {\bf Sensitivity at the LHC.}
To enhance the sensitivity to the Type~II LNV window, we propose the following search,
based on Monte Carlo simulations of the $\ell^\pm \ell^{\prime \pm} + \text{jets}$ 
final states for the signal and backgrounds at the LHC.
We focus on three possible electron and muon final states: $e^{\pm}e^{\pm}$, $\mu^{\pm}\mu^{\pm}$ 
and the $e^{\pm} \mu^{\pm}$ LFV channel.
 
We use the \texttt{Feynrules}~\cite{Alloul:2013bka} implementation of the model 
file~\cite{Fuks:2019clu} for the generation of signal events in 
the~\texttt{MadGraph5}~\cite{Alwall:2014hca} framework at LO.
\texttt{Pythia8}~\cite{Sjostrand:2007gs} is used for showering and hadronisation and the 
\texttt{Delphes}~\cite{deFavereau:2013fsa} library for fast detector simulation, using the
default detector cards.
We use~\texttt{MadAnalysis5}~\cite{Conte:2012fm} for designing and implementing the 
following cuts.
Select events with at least two same-sign leptons $\ell^{\pm} \ell^{\prime\pm}$, 
$\ell,\ell' = e, \mu$, and at least two jets, defined with the anti-$k_T$ algorithm using 
$\Delta R = 0.3$ and $p_{Tj\min} = 20 \text{ GeV}$.
Impose cuts of $p_{T \ell} > 50~\text{GeV}$ and $p_{T j} > 50~\text{GeV}$ on the leading
lepton and jet.
Accept events within the narrow di-lepton invariant mass peak, 
$m_{\ell \ell} \in [0.9, 1.1] \,m_{\Delta^{++}}$, reject those with $\Delta R_{\ell \ell} > \pi$.
Events with the invariant mass $m_{j_1 j_2} > 1.1\, m_{\Delta^{++}}$ for the two leading
jets are omitted.
\begin{figure}
  \centering
  \includegraphics[width=0.905\columnwidth]{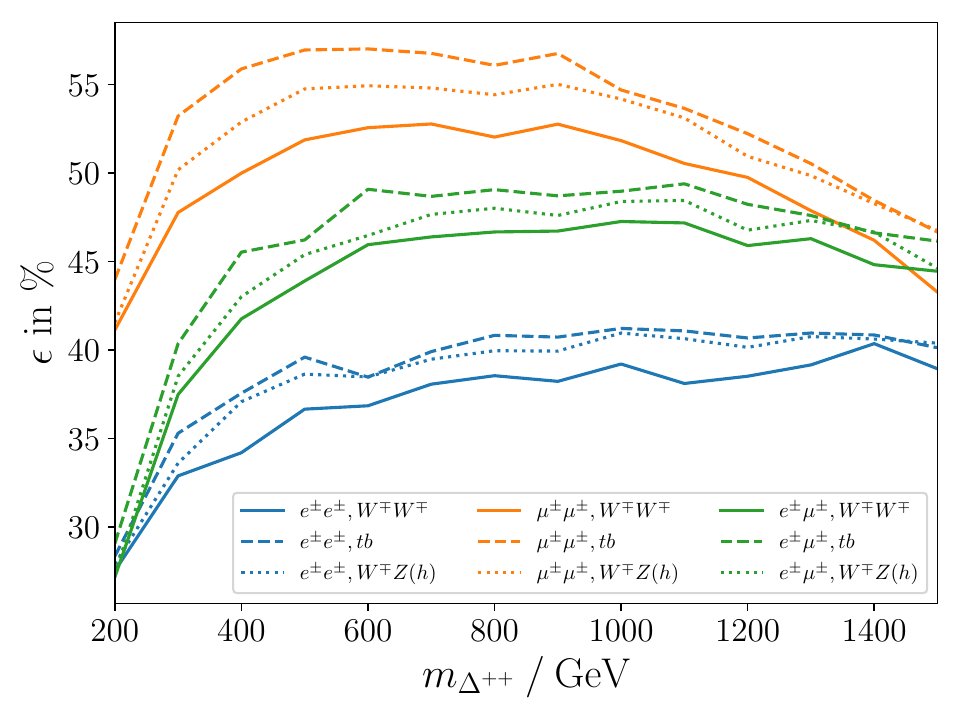}%
  \vspace*{-2ex}%
  \caption{Efficiencies after selection cuts for the signal events in the 
  $\ell^{\pm} \ell^{\pm} W^{\mp} W^{\mp}, \ell^{\pm} \ell^{\pm} W^{\mp} Z(h)$ and 
  $\ell^{\pm}\ell^{\pm} tb$ channels.}
  \label{figLNVWinEff}
\end{figure}

\begin{figure*}[!ht]
  \centering
  \mbox{\hspace{-2.5mm}\includegraphics[width=0.34\textwidth]{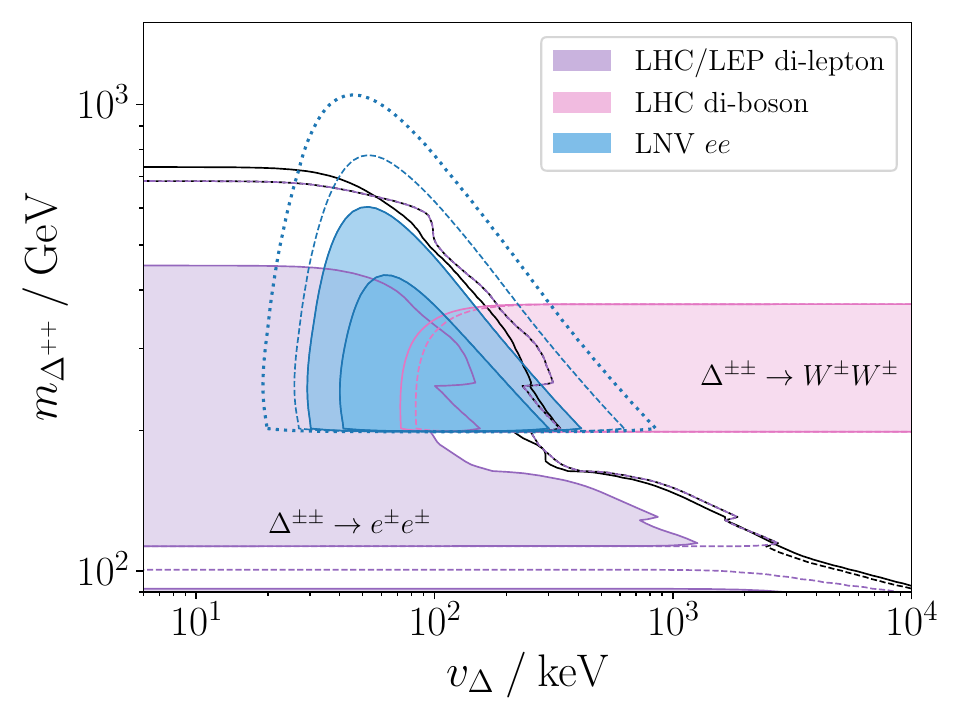}
  \hspace{-2.5mm}\includegraphics[width=0.34\textwidth]{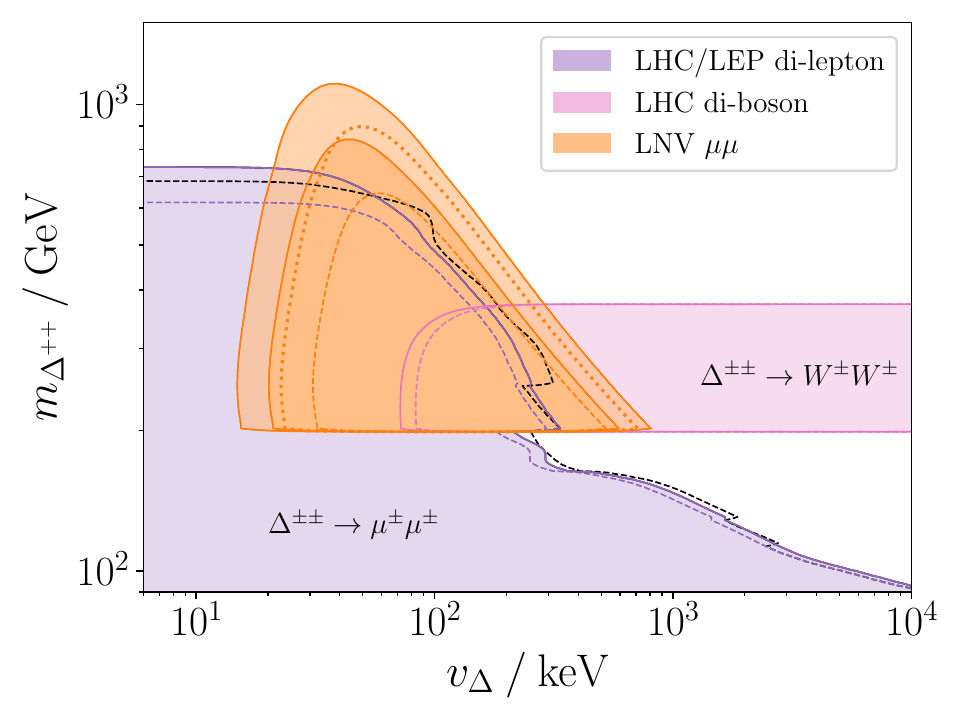}
  \hspace{-2.5mm}\includegraphics[width=0.34\textwidth]{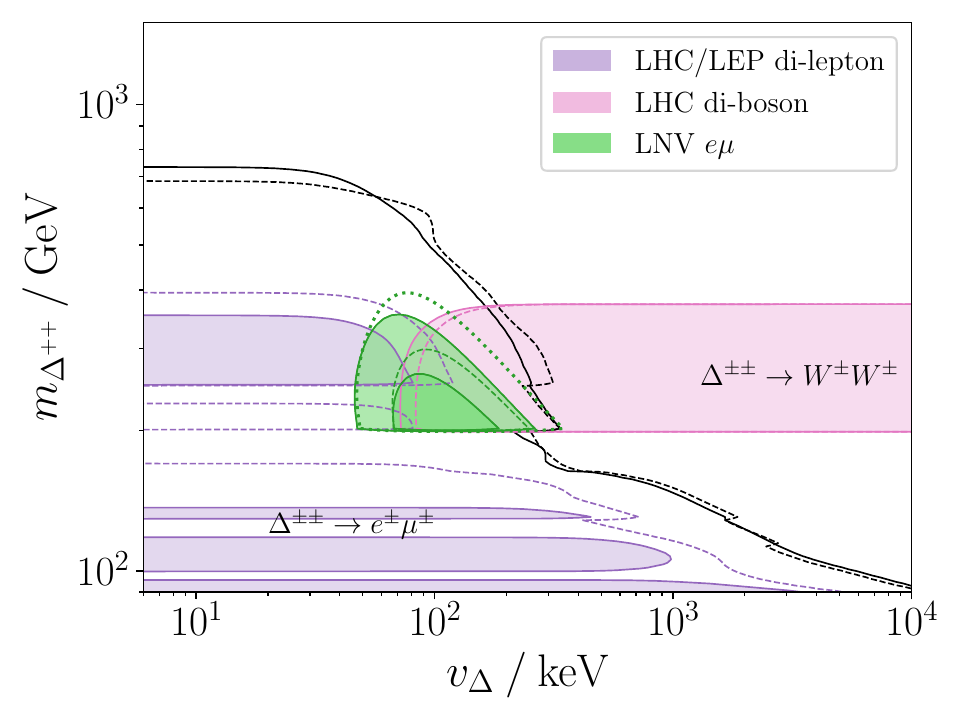}}%
  \vspace*{-2ex}%
  \caption{
  The LNV window sensitivity at $\sqrt{s} =  14$~TeV for pair and associated production of 
  $\Delta^{++}$ with $m_{\nu_{\text{min}}} = 0.01~\text{eV}$.
  The left, center and right panels correspond to the combined $e^{\pm}e^{\pm}4j$, 
  $\mu^{\pm} \mu^{\pm}4j$ and $e^\pm\mu^\pm 4j$ channels.
  The inner (outer) regions correspond to $\mathcal L = 300 \text{ fb}^{-1}$ 
  (3000~$\text{fb}^{-1}$),  solid lines show the sensitivity at $2\,\sigma$ for NO.
  The dashed (dotted) lines denote the sensitivity for IO at $300~\mathrm{fb}^{-1}$
  ($3000~\mathrm{fb}^{-1}$).
  The black solid (dashed) lines denote the strongest bound from exclusive flavour 
  channels for NO(IO).
  The purple region shows the exclusion from the doubly charged searches in the di-lepton 
  channel(s)~\cite{ATLAS:2017xqs,
  CMS:2012dun, L3:2003zst, ATLAS:2014vih, OPAL:2001luy, CMS-PAS-HIG-16-036, ATLAS:2014kca}.
  The rose region denotes the $\Delta^{\pm\pm}\rightarrow W^{\pm}W^{\pm}$ and 
  $\Delta^{\pm\pm}, \Delta^{\mp}\to W^\pm W^\pm \, , W^\mp Z$ 
  searches~\cite{ATLAS:2021jol, ATLAS:2018ceg}.
  }
  \label{figLNVWinSens}
\end{figure*}
FIG.~\ref{figLNVWinEff} shows the signal efficiencies, i.e.\ ratios between the
number of events after applying the cuts and the initial number of simulated events with 
hadronically decaying $V = W^{\pm}, Z$.

The main sources of background are $V+012js$, $VV + 012js$, $t\overline{t} + 012js$. 
All simulations were performed with \texttt{aMC@NLO} at LO with up to two matched jets 
and rescaled to the NLO cross-sections.

The estimated sensitivities $\mathcal S$, used to establish upper limits for the LNV channels,
are obtained with $\mathcal S^2 = \sum_i s_i^2/(s_i + b_i)$, where $s_i (b_i)$ is the 
expected number of signal(background) events in each region $i$ after cuts.
They are shown in FIG.~\ref{figLNVWinSens}, 
together with the exclusion regions of current searches by ATLAS, CMS, L3 and 
OPAL~\cite{ATLAS:2017xqs, CMS:2012dun, L3:2003zst, ATLAS:2014vih, OPAL:2001luy, CMS-PAS-HIG-16-036,
ATLAS:2014kca, ATLAS:2021jol, ATLAS:2018ceg} in purple.
We omit the inclusive ATLAS search~\cite{ATLAS:2022pbd}
where all flavour channels were combined into a single sensitivity.
The bounds from current searches in the leptonic channels (purple) in FIG.~\ref{figLNVWinSens}
disappear at around $v_{\Delta} \sim 100 \text{ keV}$, while for $v_{\Delta} > 200 \text{ keV}$, 
the $W^{\pm}W^{\pm}$ channel takes over, with a weaker exclusion of 
$m_{\Delta^{++}} \gtrsim 350 \text{ GeV}$ (rose).

In the intermediate $v_\Delta$ range, both leptonic and gauge searches are weakened 
and the opportunity for \emph{genuine} LNV signals emerges.
Sensitivity estimates reveal encouraging prospects justified by the extent 
of the LNV windows in FIG.~\ref{figLNVWinSens}.
They cover a large portion of parameter space, above the limits of existing searches 
and spanning orders of magnitude in $v_\Delta$.

\medskip\noindent {\bf Impact of mass splittings.}

\begin{figure}[t]
  \centering
  \includegraphics[width=\columnwidth]{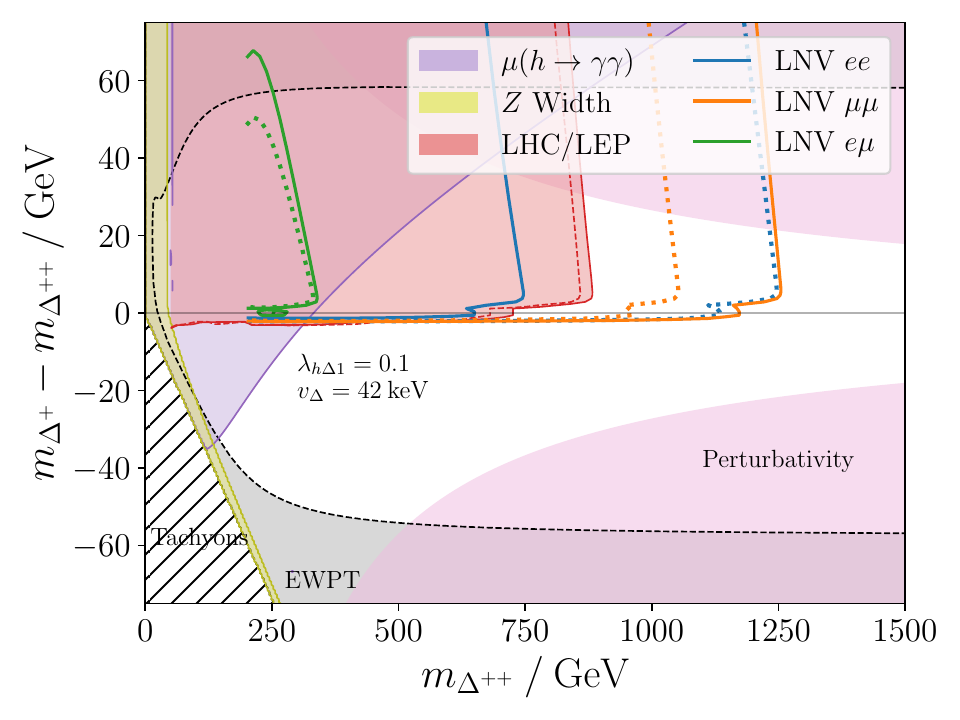}%
  \vspace*{-1.5ex}%
  \caption{
  Future sensitivity of the LHC to the LNV window for NO~(IO) in solid (dotted) lines, 
  with $\mathcal L = 3000~\mathrm{fb}^{-1}$. 
  Blue, orange and green lines denote the $e^{\pm}e^{\pm}$, $\mu^{\pm}\mu^{\pm}$ and 
  $e^{\pm}\mu^{\pm}$ channels.
  The grey area shows the EWPT constraints, the pink region excludes
  $\lambda_{h\Delta2}>\sqrt{4\pi}$, the light green region the $\Gamma_Z^\text{tot}$, 
  the purple region $h\to\gamma\gamma$ and the red region current direct searches as in FIG.~\ref{figLNVWinSens}.
  The hatched area is forbidden by the sum rule~\eqref{eqn:sumrule}.}
  \label{figLNVWinDeltam}
\end{figure}

The scalar potential contains two bi-quadratic terms that couple
$\Delta_L$ to the Higgs doublet $\varphi$,
\begin{align}
  \lambda_{h \Delta 1} \, \varphi^\dagger \varphi \,\text{Tr}\big[ \Delta_L^\dagger \Delta_L \big]
  + \lambda_{h \Delta 2} \, \text{Tr}\big[ \varphi \varphi^\dagger \Delta_L \Delta_L^\dagger \big] \, .
\end{align}
The $\lambda_{h \Delta 1}$ gives a universal mass shift to all triplet scalar components 
and is limited by $h \to \gamma \gamma$, while the $\lambda_{h \Delta 2}$ term splits them, 
such that
\begin{align} \label{eqn:sumrule}
  m_{\Delta^0}^2 - m_{\Delta^+}^2 = m_{\Delta^+}^2 - m_{\Delta^{++}}^2 = 
  \frac{\lambda_{h \Delta 2} v^2}{4} \, ,
\end{align}
up to $\mathcal O(v_\Delta/v)^2$.
The perturbativity bound $\lambda_{h \Delta 2} \lesssim \sqrt{4 \pi}$ limits the size
of the mass splitting, especially for larger $m_{\Delta^{++}}$. 
For smaller values of $m_{\Delta^{++}}$, the EWPT oblique parameters~\cite{Peskin:1991sw, 
Lavoura:1993nq, Melfo:2011nx, Chun:2012jw, Haller:2018nnx} further constrain the splitting, as 
shown in FIG.~\ref{figLNVWinDeltam}, see~\cite{Lavoura:1993nq,Cheng:2022hbo} for details.
As realized in~\cite{Melfo:2011nx}, mass splittings trigger three-body 
cascades via off-shell $V$, such as $\Delta^{++} \to \Delta^+ f \bar f$, approximated by
\begin{align} \label{eqGamCasc}
  \Gamma_{\Delta^{++} \to \Delta^+ f \bar f} & \simeq 
  \frac{3 \alpha_2^2}{5 \pi} \frac{\Delta m^5}{M_W^4} \, ,
\end{align}
with $\Delta m = m_{\Delta^{++}} - m_{\Delta^+}$. 
Two-body cascades require a large mass splitting $\Delta m > M_{W,Z}$ and are disfavoured 
by EWPT; only off-shell $V$s with soft leptons and jets are relevant.

Cascade decays modify the number of LNV events coming from the 
pair or associated production of $\Delta_L$. 
Additional sources of same-sign charged leptons are present when $\Delta^{++}$ is the lightest, 
coming from the production and cascade decays of $\Delta^{+,\,0}$.
When $\Delta^{++}$ is the heaviest, the $\Delta^{\pm\pm}\rightarrow\ell^{\pm}\ell^{\prime\pm}$ 
final state is suppressed by cascades to $\Delta^{+,\,0}$, whose decay products fail to pass 
the invariant mass cut on $m_{\ell^{\pm} \ell^{\prime\pm}}$.
The resulting sensitivity, together with current bounds from direct searches and the indirect 
constraints outlined above, is shown in FIG.~\ref{figLNVWinDeltam} (see also cascades sensitivities 
at LHC~\cite{Ashanujjaman:2021txz, Ashanujjaman:2023tlj} and $e^+ e^-$ 
colliders~\cite{Ashanujjaman:2022tdn}).

\medskip\noindent {\bf Conclusions and outlook.}
We investigate the prospect of observing LNV signals in the minimal Type~II seesaw and
delineate the region of parameter space where the sensitivity of current LNC searches 
diminishes and a genuine LNV signal becomes observable.
Our selection criteria strongly reduce the known SM backgrounds and enhances the 
sensitivity, and reveal encouraging prospects for searches at hadron colliders, filling the 
remaining gap of current searches.

The $\Delta_L$ also appears in the LRSM~\cite{Mohapatra:1980yp}, where the mass splittings are 
large if $M_{W_R}$ is fairly light~\cite{Maiezza:2016bzp}, leading
to a lower bound for $m_{\Delta^{++}}$ at around 1~TeV if $W_R$ is accessible at the LHC, 
see also~\cite{Gluza:2020qrt}.
The LRSM seesaw extends~\eqref{eqMnu} with another Type~I source, spoiling the 
relation between $Y_\Delta$ and $v_\Delta$ and modifying the flavour structure of 
final-state leptons.
This shifts the location of the LNV window, whereas the kinematics and estimated efficiencies 
still apply.
Future outlook for enhancing the sensitivity of the LNV window includes the $\tau$ final states, 
leptonic decays of $W^{\pm}$ and $Z$, and $\Delta^0, \chi_\Delta$ contributions.

Interest in the LNV window may be spurred by cosmology, where Type~II can play an important 
role in phase transitions~\cite{Zhou:2022mlz, Ghosh:2022fzp} and leptogenesis.
While the standard leptogenesis scenario with a single $\Delta_L$ fails~\cite{Hambye:2003ka}, 
a successful variant was claimed in~\cite{Barrie:2021mwi, Barrie:2022cub}.
The washout effects are strongest precisely within the LNV window~\cite{Blanchet:2008zg}.

\section*{Acknowledgements}
MN would like to thank Yue Zhang and Richard Ruiz for discussions.
PDB, JK and MN are supported by the Slovenian Research Agency under the research core funding 
No. P1-0035 and in part by the research grants J1-3013 and N1-0253.
%
%
MN is grateful to the Mainz Institute for Theoretical Physics (MITP) of the DFG Cluster of
Excellence PRISMA+ (Project ID 33083149), for its hospitality and its partial support during
the course of this work.
PDB, JK, MN and FN thank the CERN Theory group for hospitality during which a part of the 
work was being done.
MN thanks the Carleton University for hospitality and support during the completion of 
the paper.
JCV would like to thank Goran Senjanovi\'c and Michael Ramsey-Musolf for encouragement and 
enlightening discussions.
JCV was partially funded under the US Department of Energy contract
DE-SC0011095.

\def\arxiv#1[#2]{\href{http://arxiv.org/abs/#1}{[#2]}}
\def\Arxiv#1[#2]{\href{http://arxiv.org/abs/#1}{#2}}

%
%
\bibliographystyle{apsrev4-2}
\bibliography{references}

\begin{thebibliography}{89}%
\makeatletter
\providecommand \@ifxundefined [1]{%
 \@ifx{#1\undefined}
}%
\providecommand \@ifnum [1]{%
 \ifnum #1\expandafter \@firstoftwo
 \else \expandafter \@secondoftwo
 \fi
}%
\providecommand \@ifx [1]{%
 \ifx #1\expandafter \@firstoftwo
 \else \expandafter \@secondoftwo
 \fi
}%
\providecommand \natexlab [1]{#1}%
\providecommand \enquote  [1]{``#1''}%
\providecommand \bibnamefont  [1]{#1}%
\providecommand \bibfnamefont [1]{#1}%
\providecommand \citenamefont [1]{#1}%
\providecommand \href@noop [0]{\@secondoftwo}%
\providecommand \href [0]{\begingroup \@sanitize@url \@href}%
\providecommand \@href[1]{\@@startlink{#1}\@@href}%
\providecommand \@@href[1]{\endgroup#1\@@endlink}%
\providecommand \@sanitize@url [0]{\catcode `\\12\catcode `\$12\catcode
  `\&12\catcode `\#12\catcode `\^12\catcode `\_12\catcode `\%12\relax}%
\providecommand \@@startlink[1]{}%
\providecommand \@@endlink[0]{}%
\providecommand \url  [0]{\begingroup\@sanitize@url \@url }%
\providecommand \@url [1]{\endgroup\@href {#1}{\urlprefix }}%
\providecommand \urlprefix  [0]{URL }%
\providecommand \Eprint [0]{\href }%
\providecommand \doibase [0]{https://doi.org/}%
\providecommand \selectlanguage [0]{\@gobble}%
\providecommand \bibinfo  [0]{\@secondoftwo}%
\providecommand \bibfield  [0]{\@secondoftwo}%
\providecommand \translation [1]{[#1]}%
\providecommand \BibitemOpen [0]{}%
\providecommand \bibitemStop [0]{}%
\providecommand \bibitemNoStop [0]{.\EOS\space}%
\providecommand \EOS [0]{\spacefactor3000\relax}%
\providecommand \BibitemShut  [1]{\csname bibitem#1\endcsname}%
\let\auto@bib@innerbib\@empty
\bibitem [{\citenamefont {Majorana}(1937)}]{Majorana:1937vz}%
  \BibitemOpen
  \bibfield  {author} {\bibinfo {author} {\bibfnamefont {E.}~\bibnamefont
  {Majorana}},\ }\href {https://doi.org/10.1007/BF02961314} {\bibfield
  {journal} {\bibinfo  {journal} {Nuovo Cim.}\ }\textbf {\bibinfo {volume}
  {14}},\ \bibinfo {pages} {171} (\bibinfo {year} {1937})}\BibitemShut
  {NoStop}%
\bibitem [{\citenamefont {Racah}(1937)}]{Racah:1937qq}%
  \BibitemOpen
  \bibfield  {author} {\bibinfo {author} {\bibfnamefont {G.}~\bibnamefont
  {Racah}},\ }\href {https://doi.org/10.1007/BF02961321} {\bibfield  {journal}
  {\bibinfo  {journal} {Nuovo Cim.}\ }\textbf {\bibinfo {volume} {14}},\
  \bibinfo {pages} {322} (\bibinfo {year} {1937})}\BibitemShut {NoStop}%
\bibitem [{\citenamefont {Vissani}(1999)}]{Vissani:1999tu}%
  \BibitemOpen
  \bibfield  {author} {\bibinfo {author} {\bibfnamefont {F.}~\bibnamefont
  {Vissani}},\ }\href {https://doi.org/10.1088/1126-6708/1999/06/022}
  {\bibfield  {journal} {\bibinfo  {journal} {JHEP}\ }\textbf {\bibinfo
  {volume} {06}}\bibfield  {number} {\bibinfo  {number} { (022)}},\ }\Eprint
  {https://arxiv.org/abs/hep-ph/9906525} {arXiv:hep-ph/9906525} \BibitemShut
  {NoStop}%
\bibitem [{\citenamefont {Tello}\ \emph {et~al.}(2011)\citenamefont {Tello},
  \citenamefont {Nemev\v{s}ek}, \citenamefont {Nesti}, \citenamefont
  {Senjanovi\'c},\ and\ \citenamefont {Vissani}}]{Tello:2010am}%
  \BibitemOpen
  \bibfield  {author} {\bibinfo {author} {\bibfnamefont {V.}~\bibnamefont
  {Tello}}, \bibinfo {author} {\bibfnamefont {M.}~\bibnamefont {Nemev\v{s}ek}},
  \bibinfo {author} {\bibfnamefont {F.}~\bibnamefont {Nesti}}, \bibinfo
  {author} {\bibfnamefont {G.}~\bibnamefont {Senjanovi\'c}},\ and\ \bibinfo
  {author} {\bibfnamefont {F.}~\bibnamefont {Vissani}},\ }\href
  {https://doi.org/10.1103/PhysRevLett.106.151801} {\bibfield  {journal}
  {\bibinfo  {journal} {Phys. Rev. Lett.}\ }\textbf {\bibinfo {volume} {106}},\
  \bibinfo {pages} {151801} (\bibinfo {year} {2011})},\ \Eprint
  {https://arxiv.org/abs/1011.3522} {arXiv:1011.3522 [hep-ph]} \BibitemShut
  {NoStop}%
\bibitem [{\citenamefont {Dolinski}\ \emph {et~al.}(2019)\citenamefont
  {Dolinski}, \citenamefont {Poon},\ and\ \citenamefont
  {Rodejohann}}]{Dolinski:2019nrj}%
  \BibitemOpen
  \bibfield  {author} {\bibinfo {author} {\bibfnamefont {M.~J.}\ \bibnamefont
  {Dolinski}}, \bibinfo {author} {\bibfnamefont {A.~W.~P.}\ \bibnamefont
  {Poon}},\ and\ \bibinfo {author} {\bibfnamefont {W.}~\bibnamefont
  {Rodejohann}},\ }\href {https://doi.org/10.1146/annurev-nucl-101918-023407}
  {\bibfield  {journal} {\bibinfo  {journal} {Ann. Rev. Nucl. Part. Sci.}\
  }\textbf {\bibinfo {volume} {69}},\ \bibinfo {pages} {219} (\bibinfo {year}
  {2019})},\ \Eprint {https://arxiv.org/abs/1902.04097} {arXiv:1902.04097
  [nucl-ex]} \BibitemShut {NoStop}%
\bibitem [{\citenamefont {Cai}\ \emph {et~al.}(2018)\citenamefont {Cai},
  \citenamefont {Han}, \citenamefont {Li},\ and\ \citenamefont
  {Ruiz}}]{Cai:2017mow}%
  \BibitemOpen
  \bibfield  {author} {\bibinfo {author} {\bibfnamefont {Y.}~\bibnamefont
  {Cai}}, \bibinfo {author} {\bibfnamefont {T.}~\bibnamefont {Han}}, \bibinfo
  {author} {\bibfnamefont {T.}~\bibnamefont {Li}},\ and\ \bibinfo {author}
  {\bibfnamefont {R.}~\bibnamefont {Ruiz}},\ }\href
  {https://doi.org/10.3389/fphy.2018.00040} {\bibfield  {journal} {\bibinfo
  {journal} {Front. in Phys.}\ }\textbf {\bibinfo {volume} {6}},\ \bibinfo
  {pages} {40} (\bibinfo {year} {2018})},\ \Eprint
  {https://arxiv.org/abs/1711.02180} {arXiv:1711.02180 [hep-ph]} \BibitemShut
  {NoStop}%
\bibitem [{\citenamefont {Minkowski}(1977)}]{Minkowski:1977sc}%
  \BibitemOpen
  \bibfield  {author} {\bibinfo {author} {\bibfnamefont {P.}~\bibnamefont
  {Minkowski}},\ }\href {https://doi.org/10.1016/0370-2693(77)90435-X}
  {\bibfield  {journal} {\bibinfo  {journal} {Phys. Lett. B}\ }\textbf
  {\bibinfo {volume} {67}},\ \bibinfo {pages} {421} (\bibinfo {year}
  {1977})}\BibitemShut {NoStop}%
\bibitem [{\citenamefont {Mohapatra}\ and\ \citenamefont
  {Senjanovi\'c}(1980)}]{Mohapatra:1979ia}%
  \BibitemOpen
  \bibfield  {author} {\bibinfo {author} {\bibfnamefont {R.~N.}\ \bibnamefont
  {Mohapatra}}\ and\ \bibinfo {author} {\bibfnamefont {G.}~\bibnamefont
  {Senjanovi\'c}},\ }\href {https://doi.org/10.1103/PhysRevLett.44.912}
  {\bibfield  {journal} {\bibinfo  {journal} {Phys. Rev. Lett.}\ }\textbf
  {\bibinfo {volume} {44}},\ \bibinfo {pages} {912} (\bibinfo {year}
  {1980})}\BibitemShut {NoStop}%
\bibitem [{\citenamefont {Glashow}(1980)}]{Glashow:1979nm}%
  \BibitemOpen
  \bibfield  {author} {\bibinfo {author} {\bibfnamefont {S.~L.}\ \bibnamefont
  {Glashow}},\ }\href {https://doi.org/10.1007/978-1-4684-7197-7_15} {\bibfield
   {journal} {\bibinfo  {journal} {NATO Sci. Ser. B}\ }\textbf {\bibinfo
  {volume} {61}},\ \bibinfo {pages} {687} (\bibinfo {year} {1980})}\BibitemShut
  {NoStop}%
\bibitem [{\citenamefont {Gell-Mann}\ \emph {et~al.}(1979)\citenamefont
  {Gell-Mann}, \citenamefont {Ramond},\ and\ \citenamefont
  {Slansky}}]{GellMann:1980vs}%
  \BibitemOpen
  \bibfield  {author} {\bibinfo {author} {\bibfnamefont {M.}~\bibnamefont
  {Gell-Mann}}, \bibinfo {author} {\bibfnamefont {P.}~\bibnamefont {Ramond}},\
  and\ \bibinfo {author} {\bibfnamefont {R.}~\bibnamefont {Slansky}},\
  }\href@noop {} {\bibfield  {journal} {\bibinfo  {journal} {Conf. Proc. C}\
  }\textbf {\bibinfo {volume} {790927}},\ \bibinfo {pages} {315} (\bibinfo
  {year} {1979})},\ \Eprint {https://arxiv.org/abs/1306.4669} {arXiv:1306.4669
  [hep-th]} \BibitemShut {NoStop}%
\bibitem [{\citenamefont {Yanagida}(1979)}]{Yanagida:1979as}%
  \BibitemOpen
  \bibfield  {author} {\bibinfo {author} {\bibfnamefont {T.}~\bibnamefont
  {Yanagida}},\ }\href@noop {} {\bibfield  {journal} {\bibinfo  {journal}
  {Conf. Proc. C}\ }\textbf {\bibinfo {volume} {7902131}},\ \bibinfo {pages}
  {95} (\bibinfo {year} {1979})}\BibitemShut {NoStop}%
\bibitem [{\citenamefont {Pati}\ and\ \citenamefont
  {Salam}(1974)}]{Pati:1974yy}%
  \BibitemOpen
  \bibfield  {author} {\bibinfo {author} {\bibfnamefont {J.~C.}\ \bibnamefont
  {Pati}}\ and\ \bibinfo {author} {\bibfnamefont {A.}~\bibnamefont {Salam}},\
  }\href {https://doi.org/10.1103/PhysRevD.10.275} {\bibfield  {journal}
  {\bibinfo  {journal} {Phys. Rev. D}\ }\textbf {\bibinfo {volume} {10}},\
  \bibinfo {pages} {275} (\bibinfo {year} {1974})},\ \bibinfo {note} {[Erratum:
  Phys.Rev.D 11, 703--703 (1975)]}\BibitemShut {NoStop}%
\bibitem [{\citenamefont {Mohapatra}\ and\ \citenamefont
  {Pati}(1975)}]{Mohapatra:1974hk}%
  \BibitemOpen
  \bibfield  {author} {\bibinfo {author} {\bibfnamefont {R.~N.}\ \bibnamefont
  {Mohapatra}}\ and\ \bibinfo {author} {\bibfnamefont {J.~C.}\ \bibnamefont
  {Pati}},\ }\href {https://doi.org/10.1103/PhysRevD.11.566} {\bibfield
  {journal} {\bibinfo  {journal} {Phys. Rev. D}\ }\textbf {\bibinfo {volume}
  {11}},\ \bibinfo {pages} {566} (\bibinfo {year} {1975})}\BibitemShut
  {NoStop}%
\bibitem [{\citenamefont {Senjanovic}\ and\ \citenamefont
  {Mohapatra}(1975)}]{Senjanovic:1975rk}%
  \BibitemOpen
  \bibfield  {author} {\bibinfo {author} {\bibfnamefont {G.}~\bibnamefont
  {Senjanovic}}\ and\ \bibinfo {author} {\bibfnamefont {R.~N.}\ \bibnamefont
  {Mohapatra}},\ }\href {https://doi.org/10.1103/PhysRevD.12.1502} {\bibfield
  {journal} {\bibinfo  {journal} {Phys. Rev. D}\ }\textbf {\bibinfo {volume}
  {12}},\ \bibinfo {pages} {1502} (\bibinfo {year} {1975})}\BibitemShut
  {NoStop}%
\bibitem [{\citenamefont {Keung}\ and\ \citenamefont
  {Senjanovi\'c}(1983)}]{Keung:1983uu}%
  \BibitemOpen
  \bibfield  {author} {\bibinfo {author} {\bibfnamefont {W.-Y.}\ \bibnamefont
  {Keung}}\ and\ \bibinfo {author} {\bibfnamefont {G.}~\bibnamefont
  {Senjanovi\'c}},\ }\href {https://doi.org/10.1103/PhysRevLett.50.1427}
  {\bibfield  {journal} {\bibinfo  {journal} {Phys. Rev. Lett.}\ }\textbf
  {\bibinfo {volume} {50}},\ \bibinfo {pages} {1427} (\bibinfo {year}
  {1983})}\BibitemShut {NoStop}%
\bibitem [{\citenamefont {Gluza}\ \emph {et~al.}(2016)\citenamefont {Gluza},
  \citenamefont {Jelinski},\ and\ \citenamefont {Szafron}}]{Gluza:2016qqv}%
  \BibitemOpen
  \bibfield  {author} {\bibinfo {author} {\bibfnamefont {J.}~\bibnamefont
  {Gluza}}, \bibinfo {author} {\bibfnamefont {T.}~\bibnamefont {Jelinski}},\
  and\ \bibinfo {author} {\bibfnamefont {R.}~\bibnamefont {Szafron}},\ }\href
  {https://doi.org/10.1103/PhysRevD.93.113017} {\bibfield  {journal} {\bibinfo
  {journal} {Phys. Rev. D}\ }\textbf {\bibinfo {volume} {93}},\ \bibinfo
  {pages} {113017} (\bibinfo {year} {2016})},\ \Eprint
  {https://arxiv.org/abs/1604.01388} {arXiv:1604.01388 [hep-ph]} \BibitemShut
  {NoStop}%
\bibitem [{\citenamefont {Kersten}\ and\ \citenamefont
  {Smirnov}(2007)}]{Kersten:2007vk}%
  \BibitemOpen
  \bibfield  {author} {\bibinfo {author} {\bibfnamefont {J.}~\bibnamefont
  {Kersten}}\ and\ \bibinfo {author} {\bibfnamefont {A.~Y.}\ \bibnamefont
  {Smirnov}},\ }\href {https://doi.org/10.1103/PhysRevD.76.073005} {\bibfield
  {journal} {\bibinfo  {journal} {Phys. Rev. D}\ }\textbf {\bibinfo {volume}
  {76}},\ \bibinfo {pages} {073005} (\bibinfo {year} {2007})},\ \Eprint
  {https://arxiv.org/abs/0705.3221} {arXiv:0705.3221 [hep-ph]} \BibitemShut
  {NoStop}%
\bibitem [{\citenamefont {Drewes}\ \emph {et~al.}(2019)\citenamefont {Drewes},
  \citenamefont {Klari\'c},\ and\ \citenamefont {Klose}}]{Drewes:2019byd}%
  \BibitemOpen
  \bibfield  {author} {\bibinfo {author} {\bibfnamefont {M.}~\bibnamefont
  {Drewes}}, \bibinfo {author} {\bibfnamefont {J.}~\bibnamefont {Klari\'c}},\
  and\ \bibinfo {author} {\bibfnamefont {P.}~\bibnamefont {Klose}},\ }\href
  {https://doi.org/10.1007/JHEP11(2019)032} {\bibfield  {journal} {\bibinfo
  {journal} {JHEP}\ }\textbf {\bibinfo {volume} {11}}\bibfield  {number}
  {\bibinfo  {number} { (032)}},\ }\Eprint {https://arxiv.org/abs/1907.13034}
  {arXiv:1907.13034 [hep-ph]} \BibitemShut {NoStop}%
\bibitem [{\citenamefont {Maiezza}\ \emph {et~al.}(2015)\citenamefont
  {Maiezza}, \citenamefont {Nemev\v{s}ek},\ and\ \citenamefont
  {Nesti}}]{Maiezza:2015lza}%
  \BibitemOpen
  \bibfield  {author} {\bibinfo {author} {\bibfnamefont {A.}~\bibnamefont
  {Maiezza}}, \bibinfo {author} {\bibfnamefont {M.}~\bibnamefont
  {Nemev\v{s}ek}},\ and\ \bibinfo {author} {\bibfnamefont {F.}~\bibnamefont
  {Nesti}},\ }\href {https://doi.org/10.1103/PhysRevLett.115.081802} {\bibfield
   {journal} {\bibinfo  {journal} {Phys. Rev. Lett.}\ }\textbf {\bibinfo
  {volume} {115}},\ \bibinfo {pages} {081802} (\bibinfo {year} {2015})},\
  \Eprint {https://arxiv.org/abs/1503.06834} {arXiv:1503.06834 [hep-ph]}
  \BibitemShut {NoStop}%
\bibitem [{\citenamefont {Nemev\v{s}ek}\ \emph {et~al.}(2017)\citenamefont
  {Nemev\v{s}ek}, \citenamefont {Nesti},\ and\ \citenamefont
  {Vasquez}}]{Nemevsek:2016enw}%
  \BibitemOpen
  \bibfield  {author} {\bibinfo {author} {\bibfnamefont {M.}~\bibnamefont
  {Nemev\v{s}ek}}, \bibinfo {author} {\bibfnamefont {F.}~\bibnamefont
  {Nesti}},\ and\ \bibinfo {author} {\bibfnamefont {J.~C.}\ \bibnamefont
  {Vasquez}},\ }\href {https://doi.org/10.1007/JHEP04(2017)114} {\bibfield
  {journal} {\bibinfo  {journal} {JHEP}\ }\textbf {\bibinfo {volume}
  {04}}\bibfield  {number} {\bibinfo  {number} { (114)}},\ }\Eprint
  {https://arxiv.org/abs/1612.06840} {arXiv:1612.06840 [hep-ph]} \BibitemShut
  {NoStop}%
\bibitem [{\citenamefont {Nemev\v{s}ek}\ \emph {et~al.}(2018)\citenamefont
  {Nemev\v{s}ek}, \citenamefont {Nesti},\ and\ \citenamefont
  {Popara}}]{Nemevsek:2018bbt}%
  \BibitemOpen
  \bibfield  {author} {\bibinfo {author} {\bibfnamefont {M.}~\bibnamefont
  {Nemev\v{s}ek}}, \bibinfo {author} {\bibfnamefont {F.}~\bibnamefont
  {Nesti}},\ and\ \bibinfo {author} {\bibfnamefont {G.}~\bibnamefont
  {Popara}},\ }\href {https://doi.org/10.1103/PhysRevD.97.115018} {\bibfield
  {journal} {\bibinfo  {journal} {Phys. Rev. D}\ }\textbf {\bibinfo {volume}
  {97}},\ \bibinfo {pages} {115018} (\bibinfo {year} {2018})},\ \Eprint
  {https://arxiv.org/abs/1801.05813} {arXiv:1801.05813 [hep-ph]} \BibitemShut
  {NoStop}%
\bibitem [{\citenamefont {Nemev\v{s}ek}\ and\ \citenamefont
  {Nesti}(2023)}]{Nemevsek:2023hwx}%
  \BibitemOpen
  \bibfield  {author} {\bibinfo {author} {\bibfnamefont {M.}~\bibnamefont
  {Nemev\v{s}ek}}\ and\ \bibinfo {author} {\bibfnamefont {F.}~\bibnamefont
  {Nesti}},\ }\href {https://doi.org/10.1103/PhysRevD.108.015030} {\bibfield
  {journal} {\bibinfo  {journal} {Phys. Rev. D}\ }\textbf {\bibinfo {volume}
  {108}},\ \bibinfo {pages} {015030} (\bibinfo {year} {2023})},\ \Eprint
  {https://arxiv.org/abs/2306.12104} {arXiv:2306.12104 [hep-ph]} \BibitemShut
  {NoStop}%
\bibitem [{\citenamefont {Magg}\ and\ \citenamefont
  {Wetterich}(1980)}]{Magg:1980ut}%
  \BibitemOpen
  \bibfield  {author} {\bibinfo {author} {\bibfnamefont {M.}~\bibnamefont
  {Magg}}\ and\ \bibinfo {author} {\bibfnamefont {C.}~\bibnamefont
  {Wetterich}},\ }\href {https://doi.org/10.1016/0370-2693(80)90825-4}
  {\bibfield  {journal} {\bibinfo  {journal} {Phys. Lett. B}\ }\textbf
  {\bibinfo {volume} {94}},\ \bibinfo {pages} {61} (\bibinfo {year}
  {1980})}\BibitemShut {NoStop}%
\bibitem [{\citenamefont {Schechter}\ and\ \citenamefont
  {Valle}(1980)}]{Schechter:1980gr}%
  \BibitemOpen
  \bibfield  {author} {\bibinfo {author} {\bibfnamefont {J.}~\bibnamefont
  {Schechter}}\ and\ \bibinfo {author} {\bibfnamefont {J.~W.~F.}\ \bibnamefont
  {Valle}},\ }\href {https://doi.org/10.1103/PhysRevD.22.2227} {\bibfield
  {journal} {\bibinfo  {journal} {Phys. Rev. D}\ }\textbf {\bibinfo {volume}
  {22}},\ \bibinfo {pages} {2227} (\bibinfo {year} {1980})}\BibitemShut
  {NoStop}%
\bibitem [{\citenamefont {Cheng}\ and\ \citenamefont
  {Li}(1980)}]{Cheng:1980qt}%
  \BibitemOpen
  \bibfield  {author} {\bibinfo {author} {\bibfnamefont {T.~P.}\ \bibnamefont
  {Cheng}}\ and\ \bibinfo {author} {\bibfnamefont {L.-F.}\ \bibnamefont {Li}},\
  }\href {https://doi.org/10.1103/PhysRevD.22.2860} {\bibfield  {journal}
  {\bibinfo  {journal} {Phys. Rev. D}\ }\textbf {\bibinfo {volume} {22}},\
  \bibinfo {pages} {2860} (\bibinfo {year} {1980})}\BibitemShut {NoStop}%
\bibitem [{\citenamefont {Mohapatra}\ and\ \citenamefont
  {Senjanovi\'c}(1981)}]{Mohapatra:1980yp}%
  \BibitemOpen
  \bibfield  {author} {\bibinfo {author} {\bibfnamefont {R.~N.}\ \bibnamefont
  {Mohapatra}}\ and\ \bibinfo {author} {\bibfnamefont {G.}~\bibnamefont
  {Senjanovi\'c}},\ }\href {https://doi.org/10.1103/PhysRevD.23.165} {\bibfield
   {journal} {\bibinfo  {journal} {Phys. Rev. D}\ }\textbf {\bibinfo {volume}
  {23}},\ \bibinfo {pages} {165} (\bibinfo {year} {1981})}\BibitemShut
  {NoStop}%
\bibitem [{\citenamefont {Lazarides}\ \emph {et~al.}(1981)\citenamefont
  {Lazarides}, \citenamefont {Shafi},\ and\ \citenamefont
  {Wetterich}}]{Lazarides:1980nt}%
  \BibitemOpen
  \bibfield  {author} {\bibinfo {author} {\bibfnamefont {G.}~\bibnamefont
  {Lazarides}}, \bibinfo {author} {\bibfnamefont {Q.}~\bibnamefont {Shafi}},\
  and\ \bibinfo {author} {\bibfnamefont {C.}~\bibnamefont {Wetterich}},\ }\href
  {https://doi.org/10.1016/0550-3213(81)90354-0} {\bibfield  {journal}
  {\bibinfo  {journal} {Nucl. Phys. B}\ }\textbf {\bibinfo {volume} {181}},\
  \bibinfo {pages} {287} (\bibinfo {year} {1981})}\BibitemShut {NoStop}%
\bibitem [{\citenamefont {Han}\ \emph {et~al.}(2007)\citenamefont {Han},
  \citenamefont {Mukhopadhyaya}, \citenamefont {Si},\ and\ \citenamefont
  {Wang}}]{Han:2007bk}%
  \BibitemOpen
  \bibfield  {author} {\bibinfo {author} {\bibfnamefont {T.}~\bibnamefont
  {Han}}, \bibinfo {author} {\bibfnamefont {B.}~\bibnamefont {Mukhopadhyaya}},
  \bibinfo {author} {\bibfnamefont {Z.}~\bibnamefont {Si}},\ and\ \bibinfo
  {author} {\bibfnamefont {K.}~\bibnamefont {Wang}},\ }\href
  {https://doi.org/10.1103/PhysRevD.76.075013} {\bibfield  {journal} {\bibinfo
  {journal} {Phys. Rev. D}\ }\textbf {\bibinfo {volume} {76}},\ \bibinfo
  {pages} {075013} (\bibinfo {year} {2007})},\ \Eprint
  {https://arxiv.org/abs/0706.0441} {arXiv:0706.0441 [hep-ph]} \BibitemShut
  {NoStop}%
\bibitem [{\citenamefont {Fuks}\ \emph {et~al.}(2020)\citenamefont {Fuks},
  \citenamefont {Nemev\v{s}ek},\ and\ \citenamefont {Ruiz}}]{Fuks:2019clu}%
  \BibitemOpen
  \bibfield  {author} {\bibinfo {author} {\bibfnamefont {B.}~\bibnamefont
  {Fuks}}, \bibinfo {author} {\bibfnamefont {M.}~\bibnamefont {Nemev\v{s}ek}},\
  and\ \bibinfo {author} {\bibfnamefont {R.}~\bibnamefont {Ruiz}},\ }\href
  {https://doi.org/10.1103/PhysRevD.101.075022} {\bibfield  {journal} {\bibinfo
   {journal} {Phys. Rev. D}\ }\textbf {\bibinfo {volume} {101}},\ \bibinfo
  {pages} {075022} (\bibinfo {year} {2020})},\ \Eprint
  {https://arxiv.org/abs/1912.08975} {arXiv:1912.08975 [hep-ph]} \BibitemShut
  {NoStop}%
\bibitem [{\citenamefont {Chun}\ \emph {et~al.}(2003)\citenamefont {Chun},
  \citenamefont {Lee},\ and\ \citenamefont {Park}}]{Chun:2003ej}%
  \BibitemOpen
  \bibfield  {author} {\bibinfo {author} {\bibfnamefont {E.~J.}\ \bibnamefont
  {Chun}}, \bibinfo {author} {\bibfnamefont {K.~Y.}\ \bibnamefont {Lee}},\ and\
  \bibinfo {author} {\bibfnamefont {S.~C.}\ \bibnamefont {Park}},\ }\href
  {https://doi.org/10.1016/S0370-2693(03)00770-6} {\bibfield  {journal}
  {\bibinfo  {journal} {Phys. Lett. B}\ }\textbf {\bibinfo {volume} {566}},\
  \bibinfo {pages} {142} (\bibinfo {year} {2003})},\ \Eprint
  {https://arxiv.org/abs/hep-ph/0304069} {arXiv:hep-ph/0304069} \BibitemShut
  {NoStop}%
\bibitem [{\citenamefont {Garayoa}\ and\ \citenamefont
  {Schwetz}(2008)}]{Garayoa:2007fw}%
  \BibitemOpen
  \bibfield  {author} {\bibinfo {author} {\bibfnamefont {J.}~\bibnamefont
  {Garayoa}}\ and\ \bibinfo {author} {\bibfnamefont {T.}~\bibnamefont
  {Schwetz}},\ }\href {https://doi.org/10.1088/1126-6708/2008/03/009}
  {\bibfield  {journal} {\bibinfo  {journal} {JHEP}\ }\textbf {\bibinfo
  {volume} {03}}\bibfield  {number} {\bibinfo  {number} { (009)}},\ }\Eprint
  {https://arxiv.org/abs/0712.1453} {arXiv:0712.1453 [hep-ph]} \BibitemShut
  {NoStop}%
\bibitem [{\citenamefont {Kadastik}\ \emph {et~al.}(2008)\citenamefont
  {Kadastik}, \citenamefont {Raidal},\ and\ \citenamefont
  {Rebane}}]{Kadastik:2007yd}%
  \BibitemOpen
  \bibfield  {author} {\bibinfo {author} {\bibfnamefont {M.}~\bibnamefont
  {Kadastik}}, \bibinfo {author} {\bibfnamefont {M.}~\bibnamefont {Raidal}},\
  and\ \bibinfo {author} {\bibfnamefont {L.}~\bibnamefont {Rebane}},\ }\href
  {https://doi.org/10.1103/PhysRevD.77.115023} {\bibfield  {journal} {\bibinfo
  {journal} {Phys. Rev. D}\ }\textbf {\bibinfo {volume} {77}},\ \bibinfo
  {pages} {115023} (\bibinfo {year} {2008})},\ \Eprint
  {https://arxiv.org/abs/0712.3912} {arXiv:0712.3912 [hep-ph]} \BibitemShut
  {NoStop}%
\bibitem [{\citenamefont {Fileviez~Perez}\ \emph {et~al.}(2008)\citenamefont
  {Fileviez~Perez}, \citenamefont {Han}, \citenamefont {Huang}, \citenamefont
  {Li},\ and\ \citenamefont {Wang}}]{Perez:2008ha}%
  \BibitemOpen
  \bibfield  {author} {\bibinfo {author} {\bibfnamefont {P.}~\bibnamefont
  {Fileviez~Perez}}, \bibinfo {author} {\bibfnamefont {T.}~\bibnamefont {Han}},
  \bibinfo {author} {\bibfnamefont {G.-y.}\ \bibnamefont {Huang}}, \bibinfo
  {author} {\bibfnamefont {T.}~\bibnamefont {Li}},\ and\ \bibinfo {author}
  {\bibfnamefont {K.}~\bibnamefont {Wang}},\ }\href
  {https://doi.org/10.1103/PhysRevD.78.015018} {\bibfield  {journal} {\bibinfo
  {journal} {Phys. Rev. D}\ }\textbf {\bibinfo {volume} {78}},\ \bibinfo
  {pages} {015018} (\bibinfo {year} {2008})},\ \Eprint
  {https://arxiv.org/abs/0805.3536} {arXiv:0805.3536 [hep-ph]} \BibitemShut
  {NoStop}%
\bibitem [{\citenamefont {Das}\ \emph {et~al.}(2024)\citenamefont {Das},
  \citenamefont {Das},\ and\ \citenamefont {Okada}}]{Das:2024yvt}%
  \BibitemOpen
  \bibfield  {author} {\bibinfo {author} {\bibfnamefont {A.}~\bibnamefont
  {Das}}, \bibinfo {author} {\bibfnamefont {P.}~\bibnamefont {Das}},\ and\
  \bibinfo {author} {\bibfnamefont {N.}~\bibnamefont {Okada}},\ }\href@noop {}
  {\bibfield  {journal} {\bibinfo  {journal} {arXiv}\ } (\bibinfo {year}
  {2024})},\ \Eprint {https://arxiv.org/abs/2405.11820} {arXiv:2405.11820
  [hep-ph]} \BibitemShut {NoStop}%
\bibitem [{\citenamefont {Aaboud}\ \emph {et~al.}(2018)\citenamefont {Aaboud}
  \emph {et~al.}}]{ATLAS:2017xqs}%
  \BibitemOpen
  \bibfield  {author} {\bibinfo {author} {\bibfnamefont {M.}~\bibnamefont
  {Aaboud}} \emph {et~al.} (\bibinfo {collaboration} {ATLAS}),\ }\href
  {https://doi.org/10.1140/epjc/s10052-018-5661-z} {\bibfield  {journal}
  {\bibinfo  {journal} {Eur. Phys. J. C}\ }\textbf {\bibinfo {volume} {78}},\
  \bibinfo {pages} {199} (\bibinfo {year} {2018})},\ \Eprint
  {https://arxiv.org/abs/1710.09748} {arXiv:1710.09748 [hep-ex]} \BibitemShut
  {NoStop}%
\bibitem [{\citenamefont {Chatrchyan}\ \emph {et~al.}(2012)\citenamefont
  {Chatrchyan} \emph {et~al.}}]{CMS:2012dun}%
  \BibitemOpen
  \bibfield  {author} {\bibinfo {author} {\bibfnamefont {S.}~\bibnamefont
  {Chatrchyan}} \emph {et~al.} (\bibinfo {collaboration} {CMS}),\ }\href
  {https://doi.org/10.1140/epjc/s10052-012-2189-5} {\bibfield  {journal}
  {\bibinfo  {journal} {Eur. Phys. J. C}\ }\textbf {\bibinfo {volume} {72}},\
  \bibinfo {pages} {2189} (\bibinfo {year} {2012})},\ \Eprint
  {https://arxiv.org/abs/1207.2666} {arXiv:1207.2666 [hep-ex]} \BibitemShut
  {NoStop}%
\bibitem [{\citenamefont {Achard}\ \emph {et~al.}(2003)\citenamefont {Achard}
  \emph {et~al.}}]{L3:2003zst}%
  \BibitemOpen
  \bibfield  {author} {\bibinfo {author} {\bibfnamefont {P.}~\bibnamefont
  {Achard}} \emph {et~al.} (\bibinfo {collaboration} {L3}),\ }\href
  {https://doi.org/10.1016/j.physletb.2003.09.082} {\bibfield  {journal}
  {\bibinfo  {journal} {Phys. Lett. B}\ }\textbf {\bibinfo {volume} {576}},\
  \bibinfo {pages} {18} (\bibinfo {year} {2003})},\ \Eprint
  {https://arxiv.org/abs/hep-ex/0309076} {arXiv:hep-ex/0309076} \BibitemShut
  {NoStop}%
\bibitem [{\citenamefont {Aad}\ \emph {et~al.}(2015{\natexlab{a}})\citenamefont
  {Aad} \emph {et~al.}}]{ATLAS:2014vih}%
  \BibitemOpen
  \bibfield  {author} {\bibinfo {author} {\bibfnamefont {G.}~\bibnamefont
  {Aad}} \emph {et~al.} (\bibinfo {collaboration} {ATLAS}),\ }\href
  {https://doi.org/10.1007/JHEP08(2015)138} {\bibfield  {journal} {\bibinfo
  {journal} {JHEP}\ }\textbf {\bibinfo {volume} {08}}\bibfield  {number}
  {\bibinfo  {number} { (138)}},\ }\Eprint {https://arxiv.org/abs/1411.2921}
  {arXiv:1411.2921 [hep-ex]} \BibitemShut {NoStop}%
\bibitem [{\citenamefont {Abbiendi}\ \emph {et~al.}(2002)\citenamefont
  {Abbiendi} \emph {et~al.}}]{OPAL:2001luy}%
  \BibitemOpen
  \bibfield  {author} {\bibinfo {author} {\bibfnamefont {G.}~\bibnamefont
  {Abbiendi}} \emph {et~al.} (\bibinfo {collaboration} {OPAL}),\ }\href
  {https://doi.org/10.1016/S0370-2693(01)01474-5} {\bibfield  {journal}
  {\bibinfo  {journal} {Phys. Lett. B}\ }\textbf {\bibinfo {volume} {526}},\
  \bibinfo {pages} {221} (\bibinfo {year} {2002})},\ \Eprint
  {https://arxiv.org/abs/hep-ex/0111059} {arXiv:hep-ex/0111059} \BibitemShut
  {NoStop}%
\bibitem [{\citenamefont {CMS}(2017)}]{CMS-PAS-HIG-16-036}%
  \BibitemOpen
  \bibfield  {author} {\bibinfo {author} {\bibnamefont {CMS}} (\bibinfo
  {collaboration} {CMS}),\ }\href {https://cds.cern.ch/record/2242956} {\emph
  {\bibinfo {title} {{A search for doubly-charged Higgs boson production in
  three and four lepton final states at $\sqrt{s}=13~\mathrm{TeV}$}}}},\
  \bibinfo {type} {Tech. Rep.}\ (\bibinfo  {institution} {CERN},\ \bibinfo
  {address} {Geneva},\ \bibinfo {year} {2017})\BibitemShut {NoStop}%
\bibitem [{\citenamefont {Aad}\ \emph {et~al.}(2015{\natexlab{b}})\citenamefont
  {Aad} \emph {et~al.}}]{ATLAS:2014kca}%
  \BibitemOpen
  \bibfield  {author} {\bibinfo {author} {\bibfnamefont {G.}~\bibnamefont
  {Aad}} \emph {et~al.} (\bibinfo {collaboration} {ATLAS}),\ }\href
  {https://doi.org/10.1007/JHEP03(2015)041} {\bibfield  {journal} {\bibinfo
  {journal} {JHEP}\ }\textbf {\bibinfo {volume} {03}}\bibfield  {number}
  {\bibinfo  {number} { (041)}},\ }\Eprint {https://arxiv.org/abs/1412.0237}
  {arXiv:1412.0237 [hep-ex]} \BibitemShut {NoStop}%
\bibitem [{\citenamefont {Aad}\ \emph {et~al.}(2021)\citenamefont {Aad} \emph
  {et~al.}}]{ATLAS:2021jol}%
  \BibitemOpen
  \bibfield  {author} {\bibinfo {author} {\bibfnamefont {G.}~\bibnamefont
  {Aad}} \emph {et~al.} (\bibinfo {collaboration} {ATLAS}),\ }\href
  {https://doi.org/10.1007/JHEP06(2021)146} {\bibfield  {journal} {\bibinfo
  {journal} {JHEP}\ }\textbf {\bibinfo {volume} {06}}\bibfield  {number}
  {\bibinfo  {number} { (240)},\ \bibinfo {pages} {146}},\ }\Eprint
  {https://arxiv.org/abs/2101.11961} {arXiv:2101.11961 [hep-ex]} \BibitemShut
  {NoStop}%
\bibitem [{\citenamefont {Aaboud}\ \emph {et~al.}(2019)\citenamefont {Aaboud}
  \emph {et~al.}}]{ATLAS:2018ceg}%
  \BibitemOpen
  \bibfield  {author} {\bibinfo {author} {\bibfnamefont {M.}~\bibnamefont
  {Aaboud}} \emph {et~al.} (\bibinfo {collaboration} {ATLAS}),\ }\href
  {https://doi.org/10.1140/epjc/s10052-018-6500-y} {\bibfield  {journal}
  {\bibinfo  {journal} {Eur. Phys. J. C}\ }\textbf {\bibinfo {volume} {79}},\
  \bibinfo {pages} {58} (\bibinfo {year} {2019})},\ \Eprint
  {https://arxiv.org/abs/1808.01899} {arXiv:1808.01899 [hep-ex]} \BibitemShut
  {NoStop}%
\bibitem [{\citenamefont {Aad}\ \emph {et~al.}(2024)\citenamefont {Aad} \emph
  {et~al.}}]{ATLAS:2024itc}%
  \BibitemOpen
  \bibfield  {author} {\bibinfo {author} {\bibfnamefont {G.}~\bibnamefont
  {Aad}} \emph {et~al.} (\bibinfo {collaboration} {ATLAS}),\ }\href@noop {} {\
  (\bibinfo {year} {2024})},\ \Eprint {https://arxiv.org/abs/2405.04914}
  {arXiv:2405.04914 [hep-ex]} \BibitemShut {NoStop}%
\bibitem [{\citenamefont {Melfo}\ \emph {et~al.}(2012)\citenamefont {Melfo},
  \citenamefont {Nemev\v{s}ek}, \citenamefont {Nesti}, \citenamefont
  {Senjanovi\'c},\ and\ \citenamefont {Zhang}}]{Melfo:2011nx}%
  \BibitemOpen
  \bibfield  {author} {\bibinfo {author} {\bibfnamefont {A.}~\bibnamefont
  {Melfo}}, \bibinfo {author} {\bibfnamefont {M.}~\bibnamefont {Nemev\v{s}ek}},
  \bibinfo {author} {\bibfnamefont {F.}~\bibnamefont {Nesti}}, \bibinfo
  {author} {\bibfnamefont {G.}~\bibnamefont {Senjanovi\'c}},\ and\ \bibinfo
  {author} {\bibfnamefont {Y.}~\bibnamefont {Zhang}},\ }\href
  {https://doi.org/10.1103/PhysRevD.85.055018} {\bibfield  {journal} {\bibinfo
  {journal} {Phys. Rev. D}\ }\textbf {\bibinfo {volume} {85}},\ \bibinfo
  {pages} {055018} (\bibinfo {year} {2012})},\ \Eprint
  {https://arxiv.org/abs/1108.4416} {arXiv:1108.4416 [hep-ph]} \BibitemShut
  {NoStop}%
\bibitem [{\citenamefont {Primulando}\ \emph {et~al.}(2019)\citenamefont
  {Primulando}, \citenamefont {Julio},\ and\ \citenamefont
  {Uttayarat}}]{Primulando:2019evb}%
  \BibitemOpen
  \bibfield  {author} {\bibinfo {author} {\bibfnamefont {R.}~\bibnamefont
  {Primulando}}, \bibinfo {author} {\bibfnamefont {J.}~\bibnamefont {Julio}},\
  and\ \bibinfo {author} {\bibfnamefont {P.}~\bibnamefont {Uttayarat}},\ }\href
  {https://doi.org/10.1007/JHEP08(2019)024} {\bibfield  {journal} {\bibinfo
  {journal} {JHEP}\ }\textbf {\bibinfo {volume} {08}}\bibfield  {number}
  {\bibinfo  {number} { (024)}},\ }\Eprint {https://arxiv.org/abs/1903.02493}
  {arXiv:1903.02493 [hep-ph]} \BibitemShut {NoStop}%
\bibitem [{\citenamefont {Agrawal}\ \emph {et~al.}(2018)\citenamefont
  {Agrawal}, \citenamefont {Mitra}, \citenamefont {Niyogi}, \citenamefont
  {Shil},\ and\ \citenamefont {Spannowsky}}]{Agrawal:2018pci}%
  \BibitemOpen
  \bibfield  {author} {\bibinfo {author} {\bibfnamefont {P.}~\bibnamefont
  {Agrawal}}, \bibinfo {author} {\bibfnamefont {M.}~\bibnamefont {Mitra}},
  \bibinfo {author} {\bibfnamefont {S.}~\bibnamefont {Niyogi}}, \bibinfo
  {author} {\bibfnamefont {S.}~\bibnamefont {Shil}},\ and\ \bibinfo {author}
  {\bibfnamefont {M.}~\bibnamefont {Spannowsky}},\ }\href
  {https://doi.org/10.1103/PhysRevD.98.015024} {\bibfield  {journal} {\bibinfo
  {journal} {Phys. Rev. D}\ }\textbf {\bibinfo {volume} {98}},\ \bibinfo
  {pages} {015024} (\bibinfo {year} {2018})},\ \Eprint
  {https://arxiv.org/abs/1803.00677} {arXiv:1803.00677 [hep-ph]} \BibitemShut
  {NoStop}%
\bibitem [{\citenamefont {Dev}\ \emph {et~al.}(2019)\citenamefont {Dev},
  \citenamefont {Khan}, \citenamefont {Mitra},\ and\ \citenamefont
  {Rai}}]{Dev:2019hev}%
  \BibitemOpen
  \bibfield  {author} {\bibinfo {author} {\bibfnamefont {P.~S.~B.}\
  \bibnamefont {Dev}}, \bibinfo {author} {\bibfnamefont {S.}~\bibnamefont
  {Khan}}, \bibinfo {author} {\bibfnamefont {M.}~\bibnamefont {Mitra}},\ and\
  \bibinfo {author} {\bibfnamefont {S.~K.}\ \bibnamefont {Rai}},\ }\href
  {https://doi.org/10.1103/PhysRevD.99.115015} {\bibfield  {journal} {\bibinfo
  {journal} {Phys. Rev. D}\ }\textbf {\bibinfo {volume} {99}},\ \bibinfo
  {pages} {115015} (\bibinfo {year} {2019})},\ \Eprint
  {https://arxiv.org/abs/1903.01431} {arXiv:1903.01431 [hep-ph]} \BibitemShut
  {NoStop}%
\bibitem [{\citenamefont {Du}\ \emph {et~al.}(2019)\citenamefont {Du},
  \citenamefont {Dunbrack}, \citenamefont {Ramsey-Musolf},\ and\ \citenamefont
  {Yu}}]{Du:2018eaw}%
  \BibitemOpen
  \bibfield  {author} {\bibinfo {author} {\bibfnamefont {Y.}~\bibnamefont
  {Du}}, \bibinfo {author} {\bibfnamefont {A.}~\bibnamefont {Dunbrack}},
  \bibinfo {author} {\bibfnamefont {M.~J.}\ \bibnamefont {Ramsey-Musolf}},\
  and\ \bibinfo {author} {\bibfnamefont {J.-H.}\ \bibnamefont {Yu}},\ }\href
  {https://doi.org/10.1007/JHEP01(2019)101} {\bibfield  {journal} {\bibinfo
  {journal} {JHEP}\ }\textbf {\bibinfo {volume} {01}}\bibfield  {number}
  {\bibinfo  {number} { (101)}},\ }\Eprint {https://arxiv.org/abs/1810.09450}
  {arXiv:1810.09450 [hep-ph]} \BibitemShut {NoStop}%
\bibitem [{\citenamefont {Padhan}\ \emph {et~al.}(2020)\citenamefont {Padhan},
  \citenamefont {Das}, \citenamefont {Mitra},\ and\ \citenamefont
  {Kumar~Nayak}}]{Padhan:2019jlc}%
  \BibitemOpen
  \bibfield  {author} {\bibinfo {author} {\bibfnamefont {R.}~\bibnamefont
  {Padhan}}, \bibinfo {author} {\bibfnamefont {D.}~\bibnamefont {Das}},
  \bibinfo {author} {\bibfnamefont {M.}~\bibnamefont {Mitra}},\ and\ \bibinfo
  {author} {\bibfnamefont {A.}~\bibnamefont {Kumar~Nayak}},\ }\href
  {https://doi.org/10.1103/PhysRevD.101.075050} {\bibfield  {journal} {\bibinfo
   {journal} {Phys. Rev. D}\ }\textbf {\bibinfo {volume} {101}},\ \bibinfo
  {pages} {075050} (\bibinfo {year} {2020})},\ \Eprint
  {https://arxiv.org/abs/1909.10495} {arXiv:1909.10495 [hep-ph]} \BibitemShut
  {NoStop}%
\bibitem [{\citenamefont {Maiezza}\ \emph
  {et~al.}(2016{\natexlab{a}})\citenamefont {Maiezza}, \citenamefont
  {Nemev\v{s}ek},\ and\ \citenamefont {Nesti}}]{Maiezza:2016bqj}%
  \BibitemOpen
  \bibfield  {author} {\bibinfo {author} {\bibfnamefont {A.}~\bibnamefont
  {Maiezza}}, \bibinfo {author} {\bibfnamefont {M.}~\bibnamefont
  {Nemev\v{s}ek}},\ and\ \bibinfo {author} {\bibfnamefont {F.}~\bibnamefont
  {Nesti}},\ }\href {https://doi.org/10.1063/1.4953289} {\bibfield  {journal}
  {\bibinfo  {journal} {AIP Conf. Proc.}\ }\textbf {\bibinfo {volume} {1743}},\
  \bibinfo {pages} {030008} (\bibinfo {year} {2016}{\natexlab{a}})}\BibitemShut
  {NoStop}%
\bibitem [{\citenamefont {del \'Aguila}\ and\ \citenamefont
  {Chala}(2014)}]{delAguila:2013mia}%
  \BibitemOpen
  \bibfield  {author} {\bibinfo {author} {\bibfnamefont {F.}~\bibnamefont {del
  \'Aguila}}\ and\ \bibinfo {author} {\bibfnamefont {M.}~\bibnamefont
  {Chala}},\ }\href {https://doi.org/10.1007/JHEP03(2014)027} {\bibfield
  {journal} {\bibinfo  {journal} {JHEP}\ }\textbf {\bibinfo {volume}
  {03}}\bibfield  {number} {\bibinfo  {number} { (027)}},\ }\Eprint
  {https://arxiv.org/abs/1311.1510} {arXiv:1311.1510 [hep-ph]} \BibitemShut
  {NoStop}%
\bibitem [{\citenamefont {Babu}\ \emph {et~al.}(2024)\citenamefont {Babu},
  \citenamefont {Barman}, \citenamefont {Gon\c{c}alves},\ and\ \citenamefont
  {Ismail}}]{Babu:2022ycv}%
  \BibitemOpen
  \bibfield  {author} {\bibinfo {author} {\bibfnamefont {K.~S.}\ \bibnamefont
  {Babu}}, \bibinfo {author} {\bibfnamefont {R.~K.}\ \bibnamefont {Barman}},
  \bibinfo {author} {\bibfnamefont {D.}~\bibnamefont {Gon\c{c}alves}},\ and\
  \bibinfo {author} {\bibfnamefont {A.}~\bibnamefont {Ismail}},\ }\href
  {https://doi.org/10.1007/JHEP06(2024)132} {\bibfield  {journal} {\bibinfo
  {journal} {JHEP}\ }\textbf {\bibinfo {volume} {06}}\bibfield  {number}
  {\bibinfo  {number} { (132)}},\ }\Eprint {https://arxiv.org/abs/2212.08025}
  {arXiv:2212.08025 [hep-ph]} \BibitemShut {NoStop}%
\bibitem [{\citenamefont {Aghanim}\ \emph {et~al.}(2020)\citenamefont {Aghanim}
  \emph {et~al.}}]{planck:2018vyg}%
  \BibitemOpen
  \bibfield  {author} {\bibinfo {author} {\bibfnamefont {N.}~\bibnamefont
  {Aghanim}} \emph {et~al.} (\bibinfo {collaboration} {Planck}),\ }\href
  {https://doi.org/10.1051/0004-6361/201833910} {\bibfield  {journal} {\bibinfo
   {journal} {Astron. Astrophys.}\ }\textbf {\bibinfo {volume} {641}},\
  \bibinfo {pages} {A6} (\bibinfo {year} {2020})},\ \bibinfo {note} {[Erratum:
  Astron.Astrophys. 652, C4 (2021)]},\ \Eprint
  {https://arxiv.org/abs/1807.06209} {arXiv:1807.06209 [astro-ph.CO]}
  \BibitemShut {NoStop}%
\bibitem [{\citenamefont {Gando}\ \emph {et~al.}(2016)\citenamefont {Gando}
  \emph {et~al.}}]{KamLAND-Zen:2016pfg}%
  \BibitemOpen
  \bibfield  {author} {\bibinfo {author} {\bibfnamefont {A.}~\bibnamefont
  {Gando}} \emph {et~al.} (\bibinfo {collaboration} {KamLAND-Zen}),\ }\href
  {https://doi.org/10.1103/PhysRevLett.117.082503} {\bibfield  {journal}
  {\bibinfo  {journal} {Phys. Rev. Lett.}\ }\textbf {\bibinfo {volume} {117}},\
  \bibinfo {pages} {082503} (\bibinfo {year} {2016})},\ \bibinfo {note}
  {[Addendum: Phys.Rev.Lett. 117, 109903 (2016)]},\ \Eprint
  {https://arxiv.org/abs/1605.02889} {arXiv:1605.02889 [hep-ex]} \BibitemShut
  {NoStop}%
\bibitem [{\citenamefont {Aker}\ \emph {et~al.}(2022)\citenamefont {Aker} \emph
  {et~al.}}]{KATRIN:2021uub}%
  \BibitemOpen
  \bibfield  {author} {\bibinfo {author} {\bibfnamefont {M.}~\bibnamefont
  {Aker}} \emph {et~al.} (\bibinfo {collaboration} {KATRIN}),\ }\href
  {https://doi.org/10.1038/s41567-021-01463-1} {\bibfield  {journal} {\bibinfo
  {journal} {Nature Phys.}\ }\textbf {\bibinfo {volume} {18}},\ \bibinfo
  {pages} {160} (\bibinfo {year} {2022})},\ \Eprint
  {https://arxiv.org/abs/2105.08533} {arXiv:2105.08533 [hep-ex]} \BibitemShut
  {NoStop}%
\bibitem [{\citenamefont {Abada}\ \emph {et~al.}(2007)\citenamefont {Abada},
  \citenamefont {Biggio}, \citenamefont {Bonnet}, \citenamefont {Gavela},\ and\
  \citenamefont {Hambye}}]{Abada:2007ux}%
  \BibitemOpen
  \bibfield  {author} {\bibinfo {author} {\bibfnamefont {A.}~\bibnamefont
  {Abada}}, \bibinfo {author} {\bibfnamefont {C.}~\bibnamefont {Biggio}},
  \bibinfo {author} {\bibfnamefont {F.}~\bibnamefont {Bonnet}}, \bibinfo
  {author} {\bibfnamefont {M.~B.}\ \bibnamefont {Gavela}},\ and\ \bibinfo
  {author} {\bibfnamefont {T.}~\bibnamefont {Hambye}},\ }\href
  {https://doi.org/10.1088/1126-6708/2007/12/061} {\bibfield  {journal}
  {\bibinfo  {journal} {JHEP}\ }\textbf {\bibinfo {volume} {12}}\bibfield
  {number} {\bibinfo  {number} { (061)}},\ }\Eprint
  {https://arxiv.org/abs/0707.4058} {arXiv:0707.4058 [hep-ph]} \BibitemShut
  {NoStop}%
\bibitem [{\citenamefont {Fukuyama}\ \emph {et~al.}(2010)\citenamefont
  {Fukuyama}, \citenamefont {Sugiyama},\ and\ \citenamefont
  {Tsumura}}]{Fukuyama:2009xk}%
  \BibitemOpen
  \bibfield  {author} {\bibinfo {author} {\bibfnamefont {T.}~\bibnamefont
  {Fukuyama}}, \bibinfo {author} {\bibfnamefont {H.}~\bibnamefont {Sugiyama}},\
  and\ \bibinfo {author} {\bibfnamefont {K.}~\bibnamefont {Tsumura}},\ }\href
  {https://doi.org/10.1007/JHEP03(2010)044} {\bibfield  {journal} {\bibinfo
  {journal} {JHEP}\ }\textbf {\bibinfo {volume} {03}}\bibfield  {number}
  {\bibinfo  {number} { (044)}},\ }\Eprint {https://arxiv.org/abs/0909.4943}
  {arXiv:0909.4943 [hep-ph]} \BibitemShut {NoStop}%
\bibitem [{\citenamefont {Dev}\ \emph {et~al.}(2018)\citenamefont {Dev},
  \citenamefont {Ramsey-Musolf},\ and\ \citenamefont {Zhang}}]{Dev:2018sel}%
  \BibitemOpen
  \bibfield  {author} {\bibinfo {author} {\bibfnamefont {P.~S.~B.}\
  \bibnamefont {Dev}}, \bibinfo {author} {\bibfnamefont {M.~J.}\ \bibnamefont
  {Ramsey-Musolf}},\ and\ \bibinfo {author} {\bibfnamefont {Y.}~\bibnamefont
  {Zhang}},\ }\href {https://doi.org/10.1103/PhysRevD.98.055013} {\bibfield
  {journal} {\bibinfo  {journal} {Phys. Rev. D}\ }\textbf {\bibinfo {volume}
  {98}},\ \bibinfo {pages} {055013} (\bibinfo {year} {2018})},\ \Eprint
  {https://arxiv.org/abs/1806.08499} {arXiv:1806.08499 [hep-ph]} \BibitemShut
  {NoStop}%
\bibitem [{\citenamefont {Barrie}\ and\ \citenamefont
  {Petcov}(2023)}]{Barrie:2022ake}%
  \BibitemOpen
  \bibfield  {author} {\bibinfo {author} {\bibfnamefont {N.~D.}\ \bibnamefont
  {Barrie}}\ and\ \bibinfo {author} {\bibfnamefont {S.~T.}\ \bibnamefont
  {Petcov}},\ }\href {https://doi.org/10.1007/JHEP01(2023)001} {\bibfield
  {journal} {\bibinfo  {journal} {JHEP}\ }\textbf {\bibinfo {volume}
  {01}}\bibfield  {number} {\bibinfo  {number} { (001)}},\ }\Eprint
  {https://arxiv.org/abs/2210.02110} {arXiv:2210.02110 [hep-ph]} \BibitemShut
  {NoStop}%
\bibitem [{\citenamefont {Ardu}\ \emph {et~al.}(2023)\citenamefont {Ardu},
  \citenamefont {Davidson},\ and\ \citenamefont {Lavignac}}]{Ardu:2023yyw}%
  \BibitemOpen
  \bibfield  {author} {\bibinfo {author} {\bibfnamefont {M.}~\bibnamefont
  {Ardu}}, \bibinfo {author} {\bibfnamefont {S.}~\bibnamefont {Davidson}},\
  and\ \bibinfo {author} {\bibfnamefont {S.}~\bibnamefont {Lavignac}},\ }\href
  {https://doi.org/10.1007/JHEP11(2023)101} {\bibfield  {journal} {\bibinfo
  {journal} {JHEP}\ }\textbf {\bibinfo {volume} {11}}\bibfield  {number}
  {\bibinfo  {number} { (101)}},\ }\Eprint {https://arxiv.org/abs/2308.16897}
  {arXiv:2308.16897 [hep-ph]} \BibitemShut {NoStop}%
\bibitem [{\citenamefont {Banerjee}\ \emph {et~al.}(2024)\citenamefont
  {Banerjee}, \citenamefont {Englert},\ and\ \citenamefont
  {Naskar}}]{Banerjee:2024lsi}%
  \BibitemOpen
  \bibfield  {author} {\bibinfo {author} {\bibfnamefont {U.}~\bibnamefont
  {Banerjee}}, \bibinfo {author} {\bibfnamefont {C.}~\bibnamefont {Englert}},\
  and\ \bibinfo {author} {\bibfnamefont {W.}~\bibnamefont {Naskar}},\
  }\href@noop {} {\bibfield  {journal} {\bibinfo  {journal} {arXiv}\ }
  (\bibinfo {year} {2024})},\ \Eprint {https://arxiv.org/abs/2403.17455}
  {arXiv:2403.17455 [hep-ph]} \BibitemShut {NoStop}%
\bibitem [{\citenamefont {Bhupal~Dev}\ and\ \citenamefont
  {Zhang}(2018)}]{Dev:2018kpa}%
  \BibitemOpen
  \bibfield  {author} {\bibinfo {author} {\bibfnamefont {P.~S.}\ \bibnamefont
  {Bhupal~Dev}}\ and\ \bibinfo {author} {\bibfnamefont {Y.}~\bibnamefont
  {Zhang}},\ }\href {https://doi.org/10.1007/JHEP10(2018)199} {\bibfield
  {journal} {\bibinfo  {journal} {JHEP}\ }\textbf {\bibinfo {volume}
  {10}}\bibfield  {number} {\bibinfo  {number} { (199)}},\ }\Eprint
  {https://arxiv.org/abs/1808.00943} {arXiv:1808.00943 [hep-ph]} \BibitemShut
  {NoStop}%
\bibitem [{\citenamefont {Antusch}\ \emph {et~al.}(2019)\citenamefont
  {Antusch}, \citenamefont {Fischer}, \citenamefont {Hammad},\ and\
  \citenamefont {Scherb}}]{Antusch:2018svb}%
  \BibitemOpen
  \bibfield  {author} {\bibinfo {author} {\bibfnamefont {S.}~\bibnamefont
  {Antusch}}, \bibinfo {author} {\bibfnamefont {O.}~\bibnamefont {Fischer}},
  \bibinfo {author} {\bibfnamefont {A.}~\bibnamefont {Hammad}},\ and\ \bibinfo
  {author} {\bibfnamefont {C.}~\bibnamefont {Scherb}},\ }\href
  {https://doi.org/10.1007/JHEP02(2019)157} {\bibfield  {journal} {\bibinfo
  {journal} {JHEP}\ }\textbf {\bibinfo {volume} {02}}\bibfield  {number}
  {\bibinfo  {number} { (157)}},\ }\Eprint {https://arxiv.org/abs/1811.03476}
  {arXiv:1811.03476 [hep-ph]} \BibitemShut {NoStop}%
\bibitem [{\citenamefont {Alimena}\ \emph {et~al.}(2020)\citenamefont {Alimena}
  \emph {et~al.}}]{Alimena:2019zri}%
  \BibitemOpen
  \bibfield  {author} {\bibinfo {author} {\bibfnamefont {J.}~\bibnamefont
  {Alimena}} \emph {et~al.},\ }\href {https://doi.org/10.1088/1361-6471/ab4574}
  {\bibfield  {journal} {\bibinfo  {journal} {J. Phys. G}\ }\textbf {\bibinfo
  {volume} {47}},\ \bibinfo {pages} {090501} (\bibinfo {year} {2020})},\
  \Eprint {https://arxiv.org/abs/1903.04497} {arXiv:1903.04497 [hep-ex]}
  \BibitemShut {NoStop}%
\bibitem [{\citenamefont {Arbel\'aez}\ \emph {et~al.}(2019)\citenamefont
  {Arbel\'aez}, \citenamefont {Helo},\ and\ \citenamefont
  {Hirsch}}]{Arbelaez:2019cmj}%
  \BibitemOpen
  \bibfield  {author} {\bibinfo {author} {\bibfnamefont {C.}~\bibnamefont
  {Arbel\'aez}}, \bibinfo {author} {\bibfnamefont {J.~C.}\ \bibnamefont
  {Helo}},\ and\ \bibinfo {author} {\bibfnamefont {M.}~\bibnamefont {Hirsch}},\
  }\href {https://doi.org/10.1103/PhysRevD.100.055001} {\bibfield  {journal}
  {\bibinfo  {journal} {Phys. Rev. D}\ }\textbf {\bibinfo {volume} {100}},\
  \bibinfo {pages} {055001} (\bibinfo {year} {2019})},\ \Eprint
  {https://arxiv.org/abs/1906.03030} {arXiv:1906.03030 [hep-ph]} \BibitemShut
  {NoStop}%
\bibitem [{\citenamefont {Muhlleitner}\ and\ \citenamefont
  {Spira}(2003)}]{Muhlleitner:2003me}%
  \BibitemOpen
  \bibfield  {author} {\bibinfo {author} {\bibfnamefont {M.}~\bibnamefont
  {Muhlleitner}}\ and\ \bibinfo {author} {\bibfnamefont {M.}~\bibnamefont
  {Spira}},\ }\href {https://doi.org/10.1103/PhysRevD.68.117701} {\bibfield
  {journal} {\bibinfo  {journal} {Phys. Rev. D}\ }\textbf {\bibinfo {volume}
  {68}},\ \bibinfo {pages} {117701} (\bibinfo {year} {2003})},\ \Eprint
  {https://arxiv.org/abs/hep-ph/0305288} {arXiv:hep-ph/0305288} \BibitemShut
  {NoStop}%
\bibitem [{\citenamefont {Alloul}\ \emph {et~al.}(2014)\citenamefont {Alloul},
  \citenamefont {Christensen}, \citenamefont {Degrande}, \citenamefont {Duhr},\
  and\ \citenamefont {Fuks}}]{Alloul:2013bka}%
  \BibitemOpen
  \bibfield  {author} {\bibinfo {author} {\bibfnamefont {A.}~\bibnamefont
  {Alloul}}, \bibinfo {author} {\bibfnamefont {N.~D.}\ \bibnamefont
  {Christensen}}, \bibinfo {author} {\bibfnamefont {C.}~\bibnamefont
  {Degrande}}, \bibinfo {author} {\bibfnamefont {C.}~\bibnamefont {Duhr}},\
  and\ \bibinfo {author} {\bibfnamefont {B.}~\bibnamefont {Fuks}},\ }\href
  {https://doi.org/10.1016/j.cpc.2014.04.012} {\bibfield  {journal} {\bibinfo
  {journal} {Comput. Phys. Commun.}\ }\textbf {\bibinfo {volume} {185}},\
  \bibinfo {pages} {2250} (\bibinfo {year} {2014})},\ \Eprint
  {https://arxiv.org/abs/1310.1921} {arXiv:1310.1921 [hep-ph]} \BibitemShut
  {NoStop}%
\bibitem [{\citenamefont {Alwall}\ \emph {et~al.}(2014)\citenamefont {Alwall},
  \citenamefont {Frederix}, \citenamefont {Frixione}, \citenamefont {Hirschi},
  \citenamefont {Maltoni}, \citenamefont {Mattelaer}, \citenamefont {Shao},
  \citenamefont {Stelzer}, \citenamefont {Torrielli},\ and\ \citenamefont
  {Zaro}}]{Alwall:2014hca}%
  \BibitemOpen
  \bibfield  {author} {\bibinfo {author} {\bibfnamefont {J.}~\bibnamefont
  {Alwall}}, \bibinfo {author} {\bibfnamefont {R.}~\bibnamefont {Frederix}},
  \bibinfo {author} {\bibfnamefont {S.}~\bibnamefont {Frixione}}, \bibinfo
  {author} {\bibfnamefont {V.}~\bibnamefont {Hirschi}}, \bibinfo {author}
  {\bibfnamefont {F.}~\bibnamefont {Maltoni}}, \bibinfo {author} {\bibfnamefont
  {O.}~\bibnamefont {Mattelaer}}, \bibinfo {author} {\bibfnamefont {H.~S.}\
  \bibnamefont {Shao}}, \bibinfo {author} {\bibfnamefont {T.}~\bibnamefont
  {Stelzer}}, \bibinfo {author} {\bibfnamefont {P.}~\bibnamefont {Torrielli}},\
  and\ \bibinfo {author} {\bibfnamefont {M.}~\bibnamefont {Zaro}},\ }\href
  {https://doi.org/10.1007/JHEP07(2014)079} {\bibfield  {journal} {\bibinfo
  {journal} {JHEP}\ }\textbf {\bibinfo {volume} {07}}\bibfield  {number}
  {\bibinfo  {number} { (079)}},\ }\Eprint {https://arxiv.org/abs/1405.0301}
  {arXiv:1405.0301 [hep-ph]} \BibitemShut {NoStop}%
\bibitem [{\citenamefont {Sjostrand}\ \emph {et~al.}(2008)\citenamefont
  {Sjostrand}, \citenamefont {Mrenna},\ and\ \citenamefont
  {Skands}}]{Sjostrand:2007gs}%
  \BibitemOpen
  \bibfield  {author} {\bibinfo {author} {\bibfnamefont {T.}~\bibnamefont
  {Sjostrand}}, \bibinfo {author} {\bibfnamefont {S.}~\bibnamefont {Mrenna}},\
  and\ \bibinfo {author} {\bibfnamefont {P.~Z.}\ \bibnamefont {Skands}},\
  }\href {https://doi.org/10.1016/j.cpc.2008.01.036} {\bibfield  {journal}
  {\bibinfo  {journal} {Comput. Phys. Commun.}\ }\textbf {\bibinfo {volume}
  {178}},\ \bibinfo {pages} {852} (\bibinfo {year} {2008})},\ \Eprint
  {https://arxiv.org/abs/0710.3820} {arXiv:0710.3820 [hep-ph]} \BibitemShut
  {NoStop}%
\bibitem [{\citenamefont {de~Favereau}\ \emph {et~al.}(2014)\citenamefont
  {de~Favereau}, \citenamefont {Delaere}, \citenamefont {Demin}, \citenamefont
  {Giammanco}, \citenamefont {Lema\^\i{}tre}, \citenamefont {Mertens},\ and\
  \citenamefont {Selvaggi}}]{deFavereau:2013fsa}%
  \BibitemOpen
  \bibfield  {author} {\bibinfo {author} {\bibfnamefont {J.}~\bibnamefont
  {de~Favereau}}, \bibinfo {author} {\bibfnamefont {C.}~\bibnamefont
  {Delaere}}, \bibinfo {author} {\bibfnamefont {P.}~\bibnamefont {Demin}},
  \bibinfo {author} {\bibfnamefont {A.}~\bibnamefont {Giammanco}}, \bibinfo
  {author} {\bibfnamefont {V.}~\bibnamefont {Lema\^\i{}tre}}, \bibinfo {author}
  {\bibfnamefont {A.}~\bibnamefont {Mertens}},\ and\ \bibinfo {author}
  {\bibfnamefont {M.}~\bibnamefont {Selvaggi}} (\bibinfo {collaboration}
  {DELPHES 3}),\ }\href {https://doi.org/10.1007/JHEP02(2014)057} {\bibfield
  {journal} {\bibinfo  {journal} {JHEP}\ }\textbf {\bibinfo {volume}
  {02}}\bibfield  {number} {\bibinfo  {number} { (057)}},\ }\Eprint
  {https://arxiv.org/abs/1307.6346} {arXiv:1307.6346 [hep-ex]} \BibitemShut
  {NoStop}%
\bibitem [{\citenamefont {Conte}\ \emph {et~al.}(2013)\citenamefont {Conte},
  \citenamefont {Fuks},\ and\ \citenamefont {Serret}}]{Conte:2012fm}%
  \BibitemOpen
  \bibfield  {author} {\bibinfo {author} {\bibfnamefont {E.}~\bibnamefont
  {Conte}}, \bibinfo {author} {\bibfnamefont {B.}~\bibnamefont {Fuks}},\ and\
  \bibinfo {author} {\bibfnamefont {G.}~\bibnamefont {Serret}},\ }\href
  {https://doi.org/10.1016/j.cpc.2012.09.009} {\bibfield  {journal} {\bibinfo
  {journal} {Comput. Phys. Commun.}\ }\textbf {\bibinfo {volume} {184}},\
  \bibinfo {pages} {222} (\bibinfo {year} {2013})},\ \Eprint
  {https://arxiv.org/abs/1206.1599} {arXiv:1206.1599 [hep-ph]} \BibitemShut
  {NoStop}%
\bibitem [{\citenamefont {Aad}\ \emph {et~al.}(2023)\citenamefont {Aad} \emph
  {et~al.}}]{ATLAS:2022pbd}%
  \BibitemOpen
  \bibfield  {author} {\bibinfo {author} {\bibfnamefont {G.}~\bibnamefont
  {Aad}} \emph {et~al.} (\bibinfo {collaboration} {ATLAS}),\ }\href
  {https://doi.org/10.1140/epjc/s10052-023-11578-9} {\bibfield  {journal}
  {\bibinfo  {journal} {Eur. Phys. J. C}\ }\textbf {\bibinfo {volume} {83}},\
  \bibinfo {pages} {605} (\bibinfo {year} {2023})},\ \Eprint
  {https://arxiv.org/abs/2211.07505} {arXiv:2211.07505 [hep-ex]} \BibitemShut
  {NoStop}%
\bibitem [{\citenamefont {Peskin}\ and\ \citenamefont
  {Takeuchi}(1992)}]{Peskin:1991sw}%
  \BibitemOpen
  \bibfield  {author} {\bibinfo {author} {\bibfnamefont {M.~E.}\ \bibnamefont
  {Peskin}}\ and\ \bibinfo {author} {\bibfnamefont {T.}~\bibnamefont
  {Takeuchi}},\ }\href {https://doi.org/10.1103/PhysRevD.46.381} {\bibfield
  {journal} {\bibinfo  {journal} {Phys. Rev. D}\ }\textbf {\bibinfo {volume}
  {46}},\ \bibinfo {pages} {381} (\bibinfo {year} {1992})}\BibitemShut
  {NoStop}%
\bibitem [{\citenamefont {Lavoura}\ and\ \citenamefont
  {Li}(1994)}]{Lavoura:1993nq}%
  \BibitemOpen
  \bibfield  {author} {\bibinfo {author} {\bibfnamefont {L.}~\bibnamefont
  {Lavoura}}\ and\ \bibinfo {author} {\bibfnamefont {L.-F.}\ \bibnamefont
  {Li}},\ }\href {https://doi.org/10.1103/PhysRevD.49.1409} {\bibfield
  {journal} {\bibinfo  {journal} {Phys. Rev. D}\ }\textbf {\bibinfo {volume}
  {49}},\ \bibinfo {pages} {1409} (\bibinfo {year} {1994})},\ \Eprint
  {https://arxiv.org/abs/hep-ph/9309262} {arXiv:hep-ph/9309262} \BibitemShut
  {NoStop}%
\bibitem [{\citenamefont {Chun}\ \emph {et~al.}(2012)\citenamefont {Chun},
  \citenamefont {Lee},\ and\ \citenamefont {Sharma}}]{Chun:2012jw}%
  \BibitemOpen
  \bibfield  {author} {\bibinfo {author} {\bibfnamefont {E.~J.}\ \bibnamefont
  {Chun}}, \bibinfo {author} {\bibfnamefont {H.~M.}\ \bibnamefont {Lee}},\ and\
  \bibinfo {author} {\bibfnamefont {P.}~\bibnamefont {Sharma}},\ }\href
  {https://doi.org/10.1007/JHEP11(2012)106} {\bibfield  {journal} {\bibinfo
  {journal} {JHEP}\ }\textbf {\bibinfo {volume} {11}}\bibfield  {number}
  {\bibinfo  {number} { (106)}},\ }\Eprint {https://arxiv.org/abs/1209.1303}
  {arXiv:1209.1303 [hep-ph]} \BibitemShut {NoStop}%
\bibitem [{\citenamefont {Haller}\ \emph {et~al.}(2018)\citenamefont {Haller},
  \citenamefont {Hoecker}, \citenamefont {Kogler}, \citenamefont {M\"onig},
  \citenamefont {Peiffer},\ and\ \citenamefont {Stelzer}}]{Haller:2018nnx}%
  \BibitemOpen
  \bibfield  {author} {\bibinfo {author} {\bibfnamefont {J.}~\bibnamefont
  {Haller}}, \bibinfo {author} {\bibfnamefont {A.}~\bibnamefont {Hoecker}},
  \bibinfo {author} {\bibfnamefont {R.}~\bibnamefont {Kogler}}, \bibinfo
  {author} {\bibfnamefont {K.}~\bibnamefont {M\"onig}}, \bibinfo {author}
  {\bibfnamefont {T.}~\bibnamefont {Peiffer}},\ and\ \bibinfo {author}
  {\bibfnamefont {J.}~\bibnamefont {Stelzer}},\ }\href
  {https://doi.org/10.1140/epjc/s10052-018-6131-3} {\bibfield  {journal}
  {\bibinfo  {journal} {Eur. Phys. J. C}\ }\textbf {\bibinfo {volume} {78}},\
  \bibinfo {pages} {675} (\bibinfo {year} {2018})},\ \Eprint
  {https://arxiv.org/abs/1803.01853} {arXiv:1803.01853 [hep-ph]} \BibitemShut
  {NoStop}%
\bibitem [{\citenamefont {Cheng}\ \emph {et~al.}(2023)\citenamefont {Cheng},
  \citenamefont {He}, \citenamefont {Huang}, \citenamefont {Sun},\ and\
  \citenamefont {Xing}}]{Cheng:2022hbo}%
  \BibitemOpen
  \bibfield  {author} {\bibinfo {author} {\bibfnamefont {Y.}~\bibnamefont
  {Cheng}}, \bibinfo {author} {\bibfnamefont {X.-G.}\ \bibnamefont {He}},
  \bibinfo {author} {\bibfnamefont {F.}~\bibnamefont {Huang}}, \bibinfo
  {author} {\bibfnamefont {J.}~\bibnamefont {Sun}},\ and\ \bibinfo {author}
  {\bibfnamefont {Z.-P.}\ \bibnamefont {Xing}},\ }\href
  {https://doi.org/10.1016/j.nuclphysb.2023.116118} {\bibfield  {journal}
  {\bibinfo  {journal} {Nucl. Phys. B}\ }\textbf {\bibinfo {volume} {989}},\
  \bibinfo {pages} {116118} (\bibinfo {year} {2023})},\ \Eprint
  {https://arxiv.org/abs/2208.06760} {arXiv:2208.06760 [hep-ph]} \BibitemShut
  {NoStop}%
\bibitem [{\citenamefont {Ashanujjaman}\ and\ \citenamefont
  {Ghosh}(2022)}]{Ashanujjaman:2021txz}%
  \BibitemOpen
  \bibfield  {author} {\bibinfo {author} {\bibfnamefont {S.}~\bibnamefont
  {Ashanujjaman}}\ and\ \bibinfo {author} {\bibfnamefont {K.}~\bibnamefont
  {Ghosh}},\ }\href {https://doi.org/10.1007/JHEP03(2022)195} {\bibfield
  {journal} {\bibinfo  {journal} {JHEP}\ }\textbf {\bibinfo {volume}
  {03}}\bibfield  {number} {\bibinfo  {number} { (195)}},\ }\Eprint
  {https://arxiv.org/abs/2108.10952} {arXiv:2108.10952 [hep-ph]} \BibitemShut
  {NoStop}%
\bibitem [{\citenamefont {Ashanujjaman}\ and\ \citenamefont
  {Maharathy}(2023)}]{Ashanujjaman:2023tlj}%
  \BibitemOpen
  \bibfield  {author} {\bibinfo {author} {\bibfnamefont {S.}~\bibnamefont
  {Ashanujjaman}}\ and\ \bibinfo {author} {\bibfnamefont {S.~P.}\ \bibnamefont
  {Maharathy}},\ }\href {https://doi.org/10.1103/PhysRevD.107.115026}
  {\bibfield  {journal} {\bibinfo  {journal} {Phys. Rev. D}\ }\textbf {\bibinfo
  {volume} {107}},\ \bibinfo {pages} {115026} (\bibinfo {year} {2023})},\
  \Eprint {https://arxiv.org/abs/2305.06889} {arXiv:2305.06889 [hep-ph]}
  \BibitemShut {NoStop}%
\bibitem [{\citenamefont {Ashanujjaman}\ \emph {et~al.}(2022)\citenamefont
  {Ashanujjaman}, \citenamefont {Ghosh},\ and\ \citenamefont
  {Huitu}}]{Ashanujjaman:2022tdn}%
  \BibitemOpen
  \bibfield  {author} {\bibinfo {author} {\bibfnamefont {S.}~\bibnamefont
  {Ashanujjaman}}, \bibinfo {author} {\bibfnamefont {K.}~\bibnamefont
  {Ghosh}},\ and\ \bibinfo {author} {\bibfnamefont {K.}~\bibnamefont {Huitu}},\
  }\href {https://doi.org/10.1103/PhysRevD.106.075028} {\bibfield  {journal}
  {\bibinfo  {journal} {Phys. Rev. D}\ }\textbf {\bibinfo {volume} {106}},\
  \bibinfo {pages} {075028} (\bibinfo {year} {2022})},\ \Eprint
  {https://arxiv.org/abs/2205.14983} {arXiv:2205.14983 [hep-ph]} \BibitemShut
  {NoStop}%
\bibitem [{\citenamefont {Maiezza}\ \emph
  {et~al.}(2016{\natexlab{b}})\citenamefont {Maiezza}, \citenamefont
  {Nemev\v{s}ek},\ and\ \citenamefont {Nesti}}]{Maiezza:2016bzp}%
  \BibitemOpen
  \bibfield  {author} {\bibinfo {author} {\bibfnamefont {A.}~\bibnamefont
  {Maiezza}}, \bibinfo {author} {\bibfnamefont {M.}~\bibnamefont
  {Nemev\v{s}ek}},\ and\ \bibinfo {author} {\bibfnamefont {F.}~\bibnamefont
  {Nesti}},\ }\href {https://doi.org/10.1103/PhysRevD.94.035008} {\bibfield
  {journal} {\bibinfo  {journal} {Phys. Rev. D}\ }\textbf {\bibinfo {volume}
  {94}},\ \bibinfo {pages} {035008} (\bibinfo {year} {2016}{\natexlab{b}})},\
  \Eprint {https://arxiv.org/abs/1603.00360} {arXiv:1603.00360 [hep-ph]}
  \BibitemShut {NoStop}%
\bibitem [{\citenamefont {Gluza}\ \emph {et~al.}(2021)\citenamefont {Gluza},
  \citenamefont {Kordiaczynska},\ and\ \citenamefont
  {Srivastava}}]{Gluza:2020qrt}%
  \BibitemOpen
  \bibfield  {author} {\bibinfo {author} {\bibfnamefont {J.}~\bibnamefont
  {Gluza}}, \bibinfo {author} {\bibfnamefont {M.}~\bibnamefont
  {Kordiaczynska}},\ and\ \bibinfo {author} {\bibfnamefont {T.}~\bibnamefont
  {Srivastava}},\ }\href {https://doi.org/10.1088/1674-1137/abfe51} {\bibfield
  {journal} {\bibinfo  {journal} {Chin. Phys. C}\ }\textbf {\bibinfo {volume}
  {45}},\ \bibinfo {pages} {073113} (\bibinfo {year} {2021})},\ \Eprint
  {https://arxiv.org/abs/2006.04610} {arXiv:2006.04610 [hep-ph]} \BibitemShut
  {NoStop}%
\bibitem [{\citenamefont {Zhou}\ \emph {et~al.}(2022)\citenamefont {Zhou},
  \citenamefont {Bian},\ and\ \citenamefont {Du}}]{Zhou:2022mlz}%
  \BibitemOpen
  \bibfield  {author} {\bibinfo {author} {\bibfnamefont {R.}~\bibnamefont
  {Zhou}}, \bibinfo {author} {\bibfnamefont {L.}~\bibnamefont {Bian}},\ and\
  \bibinfo {author} {\bibfnamefont {Y.}~\bibnamefont {Du}},\ }\href
  {https://doi.org/10.1007/JHEP08(2022)205} {\bibfield  {journal} {\bibinfo
  {journal} {JHEP}\ }\textbf {\bibinfo {volume} {08}}\bibfield  {number}
  {\bibinfo  {number} { (205)}},\ }\Eprint {https://arxiv.org/abs/2203.01561}
  {arXiv:2203.01561 [hep-ph]} \BibitemShut {NoStop}%
\bibitem [{\citenamefont {Ghosh}\ \emph {et~al.}(2023)\citenamefont {Ghosh},
  \citenamefont {Ghosh},\ and\ \citenamefont {Roy}}]{Ghosh:2022fzp}%
  \BibitemOpen
  \bibfield  {author} {\bibinfo {author} {\bibfnamefont {P.}~\bibnamefont
  {Ghosh}}, \bibinfo {author} {\bibfnamefont {T.}~\bibnamefont {Ghosh}},\ and\
  \bibinfo {author} {\bibfnamefont {S.}~\bibnamefont {Roy}},\ }\href
  {https://doi.org/10.1007/JHEP10(2023)057} {\bibfield  {journal} {\bibinfo
  {journal} {JHEP}\ }\textbf {\bibinfo {volume} {10}}\bibfield  {number}
  {\bibinfo  {number} { (057)}},\ }\Eprint {https://arxiv.org/abs/2211.15640}
  {arXiv:2211.15640 [hep-ph]} \BibitemShut {NoStop}%
\bibitem [{\citenamefont {Hambye}\ and\ \citenamefont
  {Senjanovi\'c}(2004)}]{Hambye:2003ka}%
  \BibitemOpen
  \bibfield  {author} {\bibinfo {author} {\bibfnamefont {T.}~\bibnamefont
  {Hambye}}\ and\ \bibinfo {author} {\bibfnamefont {G.}~\bibnamefont
  {Senjanovi\'c}},\ }\href {https://doi.org/10.1016/j.physletb.2003.11.061}
  {\bibfield  {journal} {\bibinfo  {journal} {Phys. Lett. B}\ }\textbf
  {\bibinfo {volume} {582}},\ \bibinfo {pages} {73} (\bibinfo {year} {2004})},\
  \Eprint {https://arxiv.org/abs/hep-ph/0307237} {arXiv:hep-ph/0307237}
  \BibitemShut {NoStop}%
\bibitem [{\citenamefont {Barrie}\ \emph
  {et~al.}(2022{\natexlab{a}})\citenamefont {Barrie}, \citenamefont {Han},\
  and\ \citenamefont {Murayama}}]{Barrie:2021mwi}%
  \BibitemOpen
  \bibfield  {author} {\bibinfo {author} {\bibfnamefont {N.~D.}\ \bibnamefont
  {Barrie}}, \bibinfo {author} {\bibfnamefont {C.}~\bibnamefont {Han}},\ and\
  \bibinfo {author} {\bibfnamefont {H.}~\bibnamefont {Murayama}},\ }\href
  {https://doi.org/10.1103/PhysRevLett.128.141801} {\bibfield  {journal}
  {\bibinfo  {journal} {Phys. Rev. Lett.}\ }\textbf {\bibinfo {volume} {128}},\
  \bibinfo {pages} {141801} (\bibinfo {year} {2022}{\natexlab{a}})},\ \Eprint
  {https://arxiv.org/abs/2106.03381} {arXiv:2106.03381 [hep-ph]} \BibitemShut
  {NoStop}%
\bibitem [{\citenamefont {Barrie}\ \emph
  {et~al.}(2022{\natexlab{b}})\citenamefont {Barrie}, \citenamefont {Han},\
  and\ \citenamefont {Murayama}}]{Barrie:2022cub}%
  \BibitemOpen
  \bibfield  {author} {\bibinfo {author} {\bibfnamefont {N.~D.}\ \bibnamefont
  {Barrie}}, \bibinfo {author} {\bibfnamefont {C.}~\bibnamefont {Han}},\ and\
  \bibinfo {author} {\bibfnamefont {H.}~\bibnamefont {Murayama}},\ }\href
  {https://doi.org/10.1007/JHEP05(2022)160} {\bibfield  {journal} {\bibinfo
  {journal} {JHEP}\ }\textbf {\bibinfo {volume} {05}}\bibfield  {number}
  {\bibinfo  {number} { (160)}},\ }\Eprint {https://arxiv.org/abs/2204.08202}
  {arXiv:2204.08202 [hep-ph]} \BibitemShut {NoStop}%
\bibitem [{\citenamefont {Blanchet}\ \emph {et~al.}(2009)\citenamefont
  {Blanchet}, \citenamefont {Chacko},\ and\ \citenamefont
  {Mohapatra}}]{Blanchet:2008zg}%
  \BibitemOpen
  \bibfield  {author} {\bibinfo {author} {\bibfnamefont {S.}~\bibnamefont
  {Blanchet}}, \bibinfo {author} {\bibfnamefont {Z.}~\bibnamefont {Chacko}},\
  and\ \bibinfo {author} {\bibfnamefont {R.~N.}\ \bibnamefont {Mohapatra}},\
  }\href {https://doi.org/10.1103/PhysRevD.80.085002} {\bibfield  {journal}
  {\bibinfo  {journal} {Phys. Rev. D}\ }\textbf {\bibinfo {volume} {80}},\
  \bibinfo {pages} {085002} (\bibinfo {year} {2009})},\ \Eprint
  {https://arxiv.org/abs/0812.3837} {arXiv:0812.3837 [hep-ph]} \BibitemShut
  {NoStop}%
\end{thebibliography}%

\end{document}